\documentclass{emulateapj}
\usepackage{apjfonts}

\newcommand{\ang}{\AA}
\newcommand{\aox}{$\alpha_{\rm OX}$}
\newcommand{\chalf}{$c(\case{1}{2})$}
\newcommand{\civ}{{\sc C iv}}
\newcommand{\dsv}{$\delta_{\rm sim}^V$}
\newcommand{\dsw}{$\delta_{\rm sim}^{\rm FWHM}$}
\newcommand{\er}{$L_{\rm bol}/L_{\rm Edd}$}
\newcommand{\ewfe}{${\rm EW_{Fe}}$}
\newcommand{\feii}{Fe {\sc ii}}
\newcommand{\fwhmhb}{${\rm FWHM(H\beta)}$}
\newcommand{\fwhmhbb}{${\rm FWHM(H\beta_{\rm BC})}$}
\newcommand{\fwhmfe}{${\rm FWHM}_{\rm Fe}$}
\newcommand{\fwhmmg}{FWHM(Mg {\sc ii})}
\newcommand{\ergs}{${\rm ergs~s^{-1}}$}
\newcommand{\ergsa}{${\rm ergs~s^{-1}~\AA^{-1}}$}
\newcommand{\ergsh}{${\rm ergs~s^{-1}~Hz^{-1}}$}
\newcommand{\hb}{H$\beta$}
\newcommand{\hbb}{${\rm H}\beta_{\rm BC}$}
\newcommand{\hbn}{${\rm H}\beta_{\rm NC}$}
\newcommand{\kms}{$\rm km~s^{-1}$}
\newcommand{\ks}{$\chi^2$}
\newcommand{\lbol}{$L_{\rm bol}$}
\newcommand{\mbh}{$M_{\rm BH}$}
\newcommand{\mgii}{Mg {\sc ii}}

\newcommand{\myemail}{chenhu@bao.ac.cn}
\newcommand{\oiii}{{\sc [O iii]}}
\newcommand{\oii}{{\sc [O ii]}}
\newcommand{\rfe}{$R_{\rm Fe}$}
\newcommand{\ssv}{$\sigma_{\rm sim}^V$}
\newcommand{\sfv}{$\sigma_{\rm fit}^V$}
\newcommand{\ssw}{$\sigma_{\rm sim}^{\rm FWHM}$}
\newcommand{\sfw}{$\sigma_{\rm fit}^{\rm FWHM}$}
\newcommand{\vhbb}{$v({\rm H}\beta_{\rm BC})$}
\newcommand{\vhbn}{$v({\rm H}\beta_{\rm NC})$}
\newcommand{\vvhbb}{$V({\rm H}\beta_{\rm BC})$}
\newcommand{\vfe}{$v_{\rm Fe}$}
\newcommand{\vvfe}{$V_{\rm Fe}$}

\slugcomment{Accepted for publication in ApJ.}

\shorttitle{\feii\ Emission in Quasars}
\shortauthors{Hu et al.}

\begin{document}

\title{A Systematic Analysis of \feii\ Emission in Quasars:
Evidence for Inflow to the Central Black Hole}

\author{Chen Hu\altaffilmark{1,2,3}, Jian-Min Wang\altaffilmark{2},
        Luis C. Ho\altaffilmark{4},
        Yan-Mei Chen\altaffilmark{2,3}, Hao-Tong Zhang\altaffilmark{1}, 
	Wei-Hao Bian\altaffilmark{2,5} and Sui-Jian Xue\altaffilmark{1}}

\altaffiltext{1}{National Astronomical Observatories of China,
Chinese Academy of Sciences, Beijing 100012, China; \myemail}

\altaffiltext{2}{Key Laboratory for Particle Astrophysics,
Institute of High Energy Physics, Chinese Academy of Sciences,
Beijing 100039, China.}

\altaffiltext{3}{Graduate University of the Chinese Academy of Sciences,
Beijing 100049, China.}

\altaffiltext{4}{The Observatories of the Carnegie Institution of Washington,
813 Santa Barbara Street, Pasadena, CA 91101, USA.}

\altaffiltext{5}{Department of Physics and Institute of Theoretical Physics,
Nanjing Normal University, Nanjing 210097, China.}

\begin{abstract}
Broad \feii\ emission is a prominent feature of the optical and ultraviolet 
spectra of quasars.  We report on a systematical investigation of optical \feii\ 
emission in a large sample of 4037 $z<0.8$ quasars selected from the Sloan 
Digital Sky Survey.  We have developed and tested a detailed line-fitting 
technique, taking into account the complex continuum and narrow and broad 
emission-line spectrum.  Our primary goal is to quantify the velocity 
broadening and velocity shift of the \feii\ spectrum in order to constrain the 
location of the \feii-emitting region and its relation to the broad-line 
region.  We find that the majority of quasars show \feii\ emission that is 
redshifted, typically by $\sim 400$~ \kms\ but up to 2000~\kms, with respect to 
the systemic velocity of the narrow-line region or of the conventional 
broad-line region as traced by the \hb\ line.  Moreover, the line width of 
\feii\ is significantly narrower than that of the broad component of \hb.  We 
show that the magnitude of the \feii\ redshift correlates inversely with the 
Eddington ratio, and that there is a tendency for sources with redshifted \feii\ 
emission to show red asymmetry in the \hb\ line.  These characteristics 
strongly suggest that \feii\ originates from a location different from, and 
most likely exterior to, the region that produces most of \hb.  The \feii-emitting 
zone traces a portion of the broad-line region of intermediate velocities 
whose dynamics may be dominated by infall.
\end{abstract}


\section{Introduction}
Fe emission contributes significantly to the optical and ultraviolet (UV) 
spectra of most active galactic nuclei (AGNs), both in terms of wavelength 
coverage and flux.  The properties of the \feii-emitting clouds may provide 
important clues to the underlying physics in the broad-line region (BLR).  
First, \feii\ emission can be used to constrain the covering factor of BLR 
clouds from energy budget considerations.  Second, the ratio of the equivalent 
width (EW) of \feii\ to that of \hbb\ strongly varies with statistical 
measures of AGN correlations, such as the so-called Eigenvector 1 derived 
from principal component analysis, which is believed to be driven by some 
fundamental property such as mass accretion rate 
\citep[e.g.,][]{bg92,sul00a,sul00b,marzi01,marzi03a}.  Third, Fe abundance 
derived from \feii\ emission can be used to study the cosmological evolution 
of AGNs and possibly chemical enrichment of their hosts and environment 
\citep[e.g.,][]{wil85,wheel89,die02,die03,maio03}.  Careful measurement of the 
properties of \feii\ emission in a large sample of AGNs will clearly have a 
considerable impact on our understanding of these systems.

The origin of the optical/UV Fe emission has been hotly debated for
more than two decades. Thousands of Fe emission lines blend together to form a 
pseudo-continuum, which, when combined with Balmer continuum emission, results 
in the ``small blue bump'' around 3000 \ang\ \citep{grandi82,wil85}. Previously,
theoretical calculations of photoionized clouds in the BLR encountered
difficulties reproducing the observed strength of strong \feii\ emission,
prompting many authors to propose additional physical mechanisms
\citep{netzer83,wil85,joly87,joly91,collin88,sigut98}.  Recently, a more
sophisticated calculation by \citet{bal04} revealed that the predicted shape
and EW of the 2200--2800\AA\ \feii\ UV bump can only be made consistent with
observed values if either microturbulence of hundreds of \kms\
or another collisionally excited component is included in the model. 

Despite its significance, current observations of \feii\ emission provide poor 
constraints on its origin.  Some studies suggest that \feii\ emission
originates from the same region as the other broad emission lines.  For example,
\citet{phil77}, \citet{bg92}, \citet{laor97b}, and \citet{veron04} observed 
similar line widths and profiles for \feii\ and \hbb, and \citet{maoz93} found 
that both \feii\ emission and the Balmer continuum have comparable variation 
amplitudes. But there is a growing debate on this issue. \citet{marzi03b} 
found that \hbb\ is systematically broader than \feii\ for sources with 
\fwhmhbb\ $>$ 4000 \kms.  Recent studies of \feii\ emission variability have 
shown that \feii\ emission responds to variations in continuum flux but that 
the variability amplitude of \feii\ is not the same as that of \hb\ 
(\citealt{mv05} and references therein; \citealt{wang05}; \citealt{kuehn08}). 
In fact, the upper limit on the time lag between \feii\ emission and the 
continuum exceeds the lag of any other observed emission lines obtained 
\citep{mv05}.  This implies that \feii\ may be emitted from further out in the 
BLR than any other broad emission line. \citet{kuehn08} suggested that
\feii\ may be produced from a region between the BLR and the dust
sublimation radius. The study of \citet{matsuoka07}, 
based on measurements of the O {\sc i} and Ca {\sc ii} emission lines,  
supports the notion that \feii\ emerges from the outer portion of the BLR.
In a recent three-dimensional spectroscopic study of Mrk 493, 
\citet{popovic07} found that the \feii\ emission region is extensive and that 
the line width of \feii\ is only 1/3 of that of \hbb, leading them to suggest 
that \feii\ emission originates in an intermediate-line region.  If these 
findings can be confirmed, \feii\ emission can be a probe of the 
intermediate-line region, which may be the transition from the torus to 
the BLR and accretion disk.

Quasar emission lines often exhibit considerable velocity shifts with respect 
to each other.  However, the relative velocity of \feii\ emission with respect 
to other lines has not been well studied, especially in systematically for a 
large sample of objects (only a few measurements of individual sources have 
been published; e.g., \citealt{veron04}).  In fact, almost all broad quasar 
emission lines show blueward velocity shifts \citep{gastell82,carswell91}, 
with the exception of \mgii\ (no shift; \citealt{junk89}) and \hb\ (redward 
shift in some studies; e.g., \citealt{mcin99}).  Most previous studies simply 
{\it assumed}\ that \feii\ has no shift with respective to \oiii\ 
\citep[e.g.,][]{bg92,marzi96,mclure02,die03,greene05b,kim06,woo06} and that 
it has the same line width as the broad component of \hb\ 
\citep[e.g.,][]{netzer07,salv07}.  The goal of this study is to test this 
assumption.

The present paper presents the first detailed investigation of the velocity
shift and width of optical \feii\ emission\footnote{Unless otherwise noted, 
the \feii\ emission in this paper refers to the optical band.} 
in a large sample of quasars selected from 
the Sloan Digital Sky Survey (SDSS; \citealt{york00}).  Our primary motivation
is to determine the physical location and origin of the \feii-emitting region. 
We describe selection of the sample in \S \ref{sample} and
spectral analysis in \S \ref{fitting}.  We test the reliability of the 
measurements and the errors using Monte Carlo simulations (\S \ref{mcsim}),
and then check how significantly our method improves the spectral fit and how 
our method affects the measurements of other emission-line parameters (\S
\ref{compare}). Section \ref{result} discusses the results we obtained,
including the distribution of \feii\ emission shifts and widths,
correlations with other parameters, and also an analysis of the composite
spectra. The implications of our results are discussed in \S \ref{discuss},
with conclusing remarks given in \S \ref{summary}.  

Throughout this work, we adopt the following cosmological parameters: 
$H_0=70~{\rm km~s^{-1}}$ Mpc$^{-1}$, $\Omega_{m}=0.3$, and $\Omega_\Lambda=0.7$ 
\citep{spergel06}.

\section{Sample Selection} 
\label{sample} 
Our sample is selected from the SDSS Fifth Data Release (DR5;
\citealt{adelman07}) quasar catalog \citep{schneider07}.  We choose objects 
with redshift $z<0.8$ to ensure that the \oiii\ emission line lies within 
the SDSS spectral coverage.  Since the SDSS quasar sample is flux-limited and 
selected by broad-band colors \citep{richards02a}, care must be exercised in 
using it to study the quasar luminosity function \citep{vb05,richards06} or 
its cosmological evolution. However, this sample is adequate for the 
scientific goals of this work. 

We impose a series of selection criteria to ensure reliable measurements 
of \feii\ emission.  (1) We require a signal-to-noise ratio (S/N) $>$10 in the 
wavelength range 4430--5550 \ang, covering \hb, \oiii, and the most prominent 
features of optical \feii\ emission.  (2) We remove sources that have reduced 
\ks\ $>$ 4 in the continuum decomposition (\S \ref{contifit}). These 
sources cannot be fitted well by the present continuum model (Eq. 
(\ref{equ-conti}) in \S \ref{contidecomp}).  (3) We remove sources that have 
\ewfe\ $<$ 25 \ang;
this EW cut is determined by the simulations described in \S \ref{mcsim}.
(4) We also remove sources with \hbb\ FWHM errors $>$10\% and \oiii\
$\lambda$5007 peak velocity shift errors $>$100 \kms.  In total, the final 
sample contains 4037 sources, which is roughly 30\% of all quasars in DR5 with 
redshift $z<0.8$.

\section{Spectral Fitting}
\label{fitting}
The spectra of quasars from UV to optical wavelengths are remarkably similar to
each other.  The quasar composite spectrum \citep[e.g.,][]{vb01} is 
characterized by a featureless continuum and a plethora of broad and narrow 
emission lines. In the luminosity range of interest to us here, very little, 
if any, starlight is observed, so we can ignore the host galaxy contribution 
to the spectrum.  The validity of this simplification can be tested from our 
spectral fitting---nearly all of the sources in the sample can be well fitted 
without a host galaxy component.

The procedure of our spectral fitting algorithm is as follows.  We begin by
deredshifting the spectrum after correcting for Galactic extinction. Then, the
continuum is decomposed into three components: (1) a single power law, (2)
Balmer continuum emission (supplemented with high-order Balmer emission
lines), and (3) a pseudo-continuum due to blended Fe emission. We subtract this
continuum model to obtain a pure emission-line spectrum, which is then fitted
to derive parameters for the emission lines.  The measurement of Fe emission 
is strongly affected by the uncertainty in the determination of the continuum 
level. Since there are almost no ``pure'' continuum windows, a simultaneous fit
should be performed to decouple the Fe emission from the featureless 
continuum, rather than fitting the two independently.  [Tsuzuki et al. (2006) 
adopt an alternative approach in which they use a theoretical model to 
estimate the flux fraction of the emission lines in the continuum windows.]
The following subsections will describe each step in detail.

\subsection{Continuum Decomposition}
\label{contidecomp}
We use the $R_V$-dependent Galactic extinction law given by \citet{ccm89} and 
assume $R_V=3.1$. Eqs. (2a) and (2b) in \citet{ccm89} are used for the 
infrared band, but Eqs. (3a) and (3b) are replaced by those in 
\citet{odonnell94} for the optical band. We adopt the Galactic extinction in 
the $u$ band listed in the SDSS quasar catalog \citep{schneider07} and change 
it to the $V$ band using the relation $A_V=A_u/1.579$ \citep{schlegel98}. Then,
we deredshift the extinction-corrected spectrum using the redshift provided by 
the SDSS pipeline. This value of redshift is later refined using the line 
centroid of \oiii\ $\lambda$5007 measured after continuum decomposition 
(see \S \ref{redshift}).  

We model the continuum of the spectrum after Galactic extinction and redshift 
correction (hereinafter the corrected spectrum) using three components:
\begin{eqnarray}
  \label{equ-conti}
  F_\lambda & = & F_\lambda^{\rm PL}(F_{5100},\alpha)
  +F_\lambda^{\rm BaC}(F_{\rm BE},\tau_{\rm BE}) \nonumber\\
  & & +F_\lambda^{\rm Fe}(F_{\rm Fe}, {\rm FWHM_{\rm Fe}}, V_{\rm Fe}).
\end{eqnarray}
In total there are seven free parameters.  The first term is the featureless 
power law
\begin{equation}
  F_\lambda^{\rm PL}=F_{5100}
  \left( \frac{\lambda}{5100~{\rm \AA}}\right) ^{\alpha},
\end{equation}
where $F_{5100}$ is the flux density at 5100 \AA\ and $\alpha$ is the spectral 
index.    The second and third terms denote the Balmer continuum and Fe 
emission, respectively, which are described below.  For the Fe emission term,
$F_{\rm Fe}$ is the flux and $V_{\rm Fe}$ is the shift velocity of \feii.

\subsubsection{Balmer Continuum and High-order Balmer Lines}
\label{balconti}
Following \citet{grandi82} and \citet{die02}, the Balmer continuum produced by
a partially optically thick cloud with a uniform temperature can be expressed
by
\begin{equation}
  F_\lambda^{\rm BaC}=F_{\rm BE}B_\lambda(T_e)(1-e^{-\tau_\lambda})
\end{equation}
for wavelengths shortward of the Balmer edge ($\lambda_{\rm BE} = 3646$ \AA).  
$B_\lambda(T_e)$ is the Planck function at an electron temperature $T_e$, 
$F_{\rm BE}$ is a normalization coefficient for the flux at $\lambda_{\rm BE}$, 
and $\tau_\lambda$ is the optical depth at $\lambda$ expressed by
\begin{equation}
  \tau_\lambda=\tau_{\rm BE}\left( \frac{\lambda}{\lambda_{\rm BE}}\right),
\end{equation}
where $\tau_{\rm BE}$ is the optical depth at the Balmer edge.  There are two
free parameters, $F_{\rm BE}$ and $\tau_{\rm BE}$. Following \citet{die02}, 
we assume $T_e$ = 15,000 K.

At wavelengths $\lambda>\lambda_{\rm BE}$, blended higher-order Balmer lines 
give a smooth rise in the spectrum from $\sim$4000 \ang\ to the Balmer
edge \citep{wil85}. We treat the higher-order Balmer lines in a manner similar 
to that in \citet{die03}, with some modifications. To determine the relative 
strengths of the transitions with $7\le n\le50$, we use the line emissivities
given by \citet{storey95} for Case B, $T_e$ = 15,000 K, and $n_e=10^8~{\rm
cm^{-3}}$.  We normalize the flux of the higher-order Balmer lines to the
flux of the Balmer continuum at the edge using the results in \citet{wil85}.
(This implies that the higher-order Balmer lines also depend on $F_{\rm BE}$ 
and $\tau_{\rm BE}$.) In order to smooth the rise to the Balmer edge, we 
assume that each line has a Gaussian profile with FWHM = 8000 \kms.  In 
practice, none of the assumptions concerning the higher-order Balmer lines
actually impact our results because our fitting windows do not include this
region (see \S \ref{contifit}).

\subsubsection{Fe Emission}
\label{fefitting}
\citet{phil77} first introduced the template-fitting method to treat Fe 
emission in AGNs, using the Fe spectrum of the narrow-line Seyfert 1 (NLS1) 
galaxy I~Zw~1 ($z = 0.061$) to construct an \feii\ template.  In most 
applications of this method, the amount of velocity broadening applied 
to the template during spectral fitting is either solved as a free parameter
or is fixed (usually to the FWHM of broad \hb).  But the Fe template itself 
is not allowed to shift in velocity.  We follow essentially the same 
template-fitting, but we explicitly allow the width and shift of \feii\ to 
be free parameters.  We now describe the details of our algorithm.

Considering the redshift range of our sample, we need both the UV and optical 
Fe template for I~Zw~1. In the UV band, we adopt the Fe template of 
\citet{mv01}. Note that their template is set to zero around the \mgii\ line,
which is unphysical. However, the \mgii\ line is not the main focus of the
present work, and for the purposes of this work we do not concern ourselves 
with this complication.
In the optical, apart from the widely 
used template constructed by \citet{bg92}, some others are also available.
For example, a template of any width can be constructed from the list of Fe 
lines for I~Zw~1 given by \citet{veron04}.  We compared these two different 
templates and in the end chose the one from \citet{bg92} (kindly provided by 
T. A. Boroson) because it gives smaller reduced \ks.  However, we have 
verified that the shift of Fe emission, one of the main goals of this work, is 
actually not sensitive to the choice of template.  The velocity shifts 
measured using either of the two template are consistent with each other.

We combine the UV and optical templates to form a single template (with a gap 
from 3100 \ang\ to 3700 \ang, which has no data) and convolve it with a 
Gaussian function:
\begin{equation}
  \label{equ-fe}
  F_\lambda^{\rm Fe}=F_\lambda^{\rm I Zw 1}\ast G(F_{\rm conv}, {\rm
  FWHM_{conv}}, V_{\rm conv}),
\end{equation}
where $F_\lambda^{\rm I Zw 1}$ is the I Zw 1 Fe template, $G$ is a Gaussian 
function with flux $F_{\rm conv}$, width ${\rm FWHM_{conv}}$, and peak velocity
shift $V_{\rm conv}$. The convolution is done in logarithmic wavelength
space because $d({\rm ln}\lambda)=d\lambda/\lambda=dv/c$. The parameters in
Eq. (\ref{equ-conti}) can be calculated as follows.  The flux of the Fe
emission, $F_{\rm Fe}$, is equal to $F_{\rm conv}$ multiplied by the flux of the
template.  The shift of the Fe spectrum, \vvfe, is simply $V_{\rm conv}$. 
Finally, the FWHM of the Fe lines can be expressed as
\begin{equation}
  {\rm FWHM_{\rm Fe}}=\sqrt{{\rm FWHM_{I Zw 1}^2+FWHM_{conv}^2}}~~~.
\end{equation}

In the above algorithm, we assume that the Fe emission in the UV and optical 
have the same width and velocity shift, and that the ratio of UV Fe flux
to optical Fe flux is fixed to that of I~Zw~1. These assumptions help to 
reduce the number of free parameters, and seem appropriate given the S/N of 
the present sample.  If we split the UV/optical template following
\citet{verner04} to three major wavelength bands---UV (2000--3000 \ang), small
blue bump (3000--3500 \ang), and optical (4000--6000 \ang)---the final results
will be determined mainly by the optical Fe emission. The reasons are as 
follows. First, the small blue bump Fe emission is outside of our fitting 
windows (see \S\ref{contifit}). Second, the spectra of most of the quasars in 
the sample either do not cover or only cover a very narrow segment of the UV
Fe emission wavelength range. Third, the S/N of the UV band is lower than
the optical band. Thus, our measurements mainly trace the optical Fe
emission and are not very sensitive to UV Fe emission. 
We consider each of the three assumptions in turn.

The ratio of UV to optical Fe flux has been investigated by many authors
\citep[e.g.,][and references therein]{sigut04,verner04,bal04}. This ratio
depends on the physical parameters of the clouds, such as the hydrogen density
$n_{\rm H}$, the hydrogen-ionizing flux $\Phi_{\rm H}$, the velocity of
turbulence, and so forth. However, for physical conditions typical of quasars 
\citep[$n_{\rm H}\approx 10^{11}~{\rm cm^{-3}}$, $\Phi_{\rm H}\approx
3\times 10^{20}~{\rm cm^{-2}~s^{-1}}$;][]{ferland92}, models of Fe emission 
indicate that there are large regions of parameter space where the ratio 
is roughly constant \citep[Figs. 3 and 4. in][]{verner04}.  This suggests that 
adopting a single ratio of UV to optical Fe flux (fixed to that of I~Zw~1)
should be a reasonably good approximation.

The width of the UV Fe lines is also not necessarily equal to that of the 
optical Fe lines. Many authors fix the width of the optical Fe lines to the 
width of broad \hbb, while in the UV the width is fixed to that of \mgii\
\citep[e.g.,][]{salv07}.  This procedure implicitly assumes that the optical 
Fe lines and \hbb\ originate from the same region, and similarly that the 
UV Fe lines follow \mgii.  Empirically, however, the width of \hbb\ is 
consistent with that of \mgii\ \citep{mclure02}; the difference is only 0.05 
dex on average \citep{salv07}. Our results in \S \ref{resultwidth} also show 
this. Thus, for the present purposes, it is safe to assume one single value for
the width of the optical and UV Fe lines. We leave the width as a free 
parameter, as has been done in many previous studies 
\citep[e.g.,][]{mclure02,die03,kim06}.

The velocity shift of the UV Fe lines\footnote{In fact, the velocity shift of
each individual UV Fe line may be different. \citet{mv01} measured the
velocity of each UV Fe line and showed that their shifts can vary by as much
as $\sim$ 100 \kms.  But considering the resolution of our spectra, assuming 
a single velocity shift is safe, both for the UV and optical.} could be 
different from that of the optical lines if they arise from different 
regions. We use only the optical Fe template and fit the continuum only in
the optical band for testing. The resulting Fe shifts change
little, demonstrating that our measurements are mainly determined by the
optical Fe lines. Thus, for our goal of studying the optical Fe lines,
assuming one single shift for the UV and optical Fe lines is an adequate
approximation.

It should be noted here that because of the two reasons mentioned above (low
S/N around UV Fe lines and incomplete wavelength near), the present paper
cannot conclude whether the UV and optical \feii\ have a common origin.

\subsubsection{Multicomponent Fit}
\label{contifit}

The continuum model described by Eq. (\ref{equ-conti}) is fitted by 
minimizing the quantity
\begin{equation}
  \chi^2=\sum_i\left(\frac{y_i-y_{\rm model}}{\sigma_i}\right)^2,
\end{equation}
where $\sigma_i$ is the error of the data set $(x_i,y_i)$.  We adopt the 
Levenberg-Marquardt method \citep[chap.  15.5]{press92} to solve 
Eq. (\ref{equ-conti}), 
which is  nonlinear. We also use it in fitting the emission lines (\S
\ref{linefit}).  The fitting is performed in the following windows:
2470--2625, 2675--2755, 2855--3010, 3625--3645, 4170--4260, 4430--4770,
5080--5550, 6050--6200, and 6890--7010 \ang. These windows are devoid of strong
emission lines \citep{mv01,kim06}. The window 3625-3645 \ang\ is used to
constrain the Balmer continuum emission, because in this region there is no
strong Fe emission \citep{wil85}. The reduced \ks\ distribution has a median value of 
\ks\ $=$ 1.365. 

\begin{figure*}
  \centering
  \includegraphics[angle=-90,width=0.9\textwidth]{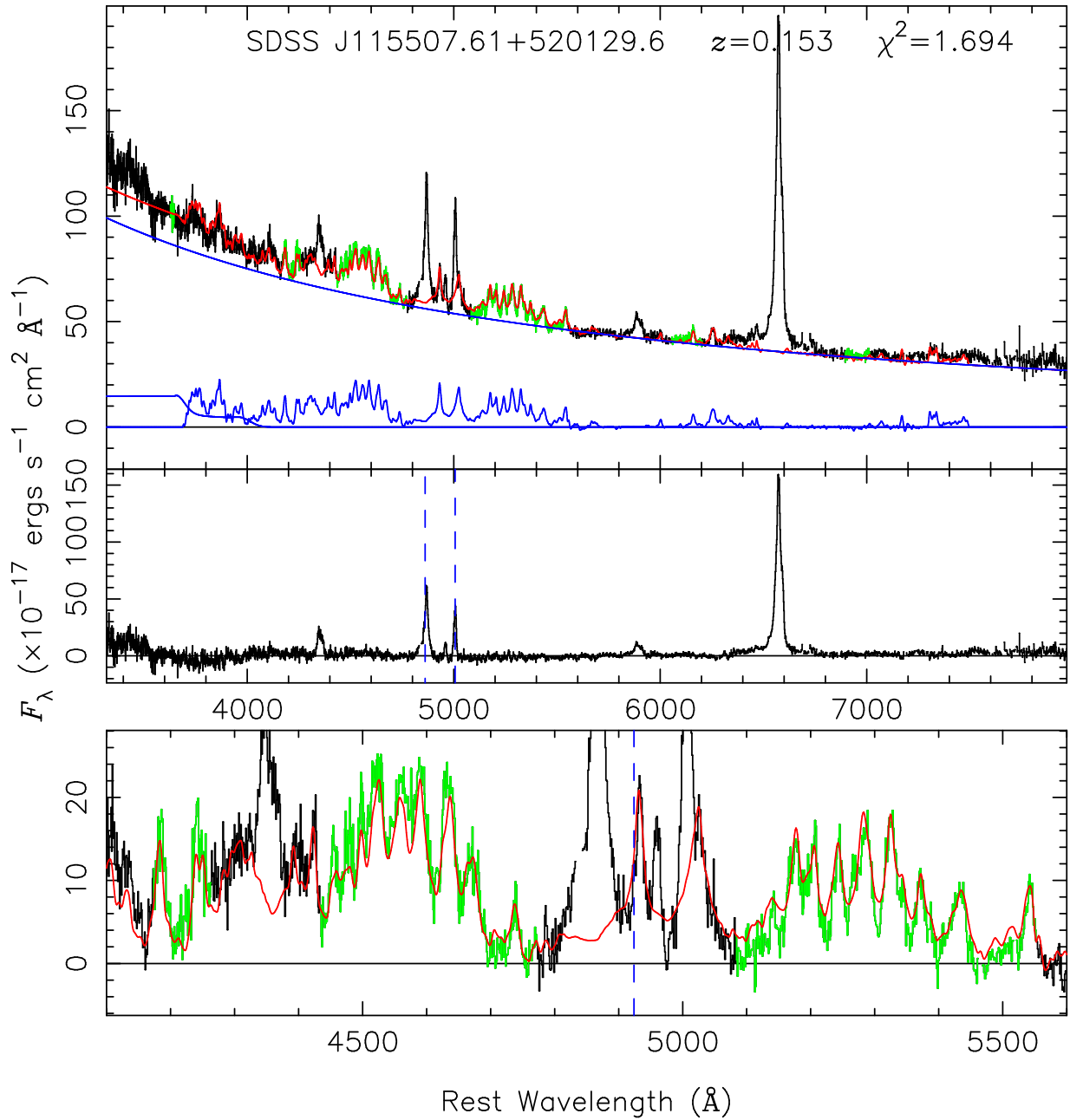}
  \caption{
  Example of continuum decomposition for SDSS J115507.61+520129.6, which has 
  narrow Fe lines. Its \vfe\ is 
  459$\pm$16 \kms. Note that our Fe model fits the two strong \feii\ lines at 
  4924 and 5018 \ang\ very well, even though these two lines are not in 
  the fitting window.  The top panel gives the spectrum after Galactic 
  extinction and redshift correction. The spectrum in the fitting window is 
  plotted in green; each component is plotted in blue; and the summed 
  continuum is plotted in red. The middle panel shows the residual, 
  pure emission-line spectrum. Two blue dashed lines mark the positions of 
  \hb\ $\lambda$4861 and \oiii\ $\lambda$5007.
  The bottom panel shows the spectrum after subtracting the power law and the
  Balmer continuum in the wavelength range 4100--5600 \ang.  The red spectrum
  is our Fe model. The blue dashed line in the bottom panel is the
  position of the peak of \feii\ $\lambda$4924 with zero velocity
  shift.  
  }
  \label{fig-conti1}
\end{figure*}

\begin{figure*}
  \centering
  \includegraphics[angle=-90,width=0.9\textwidth]{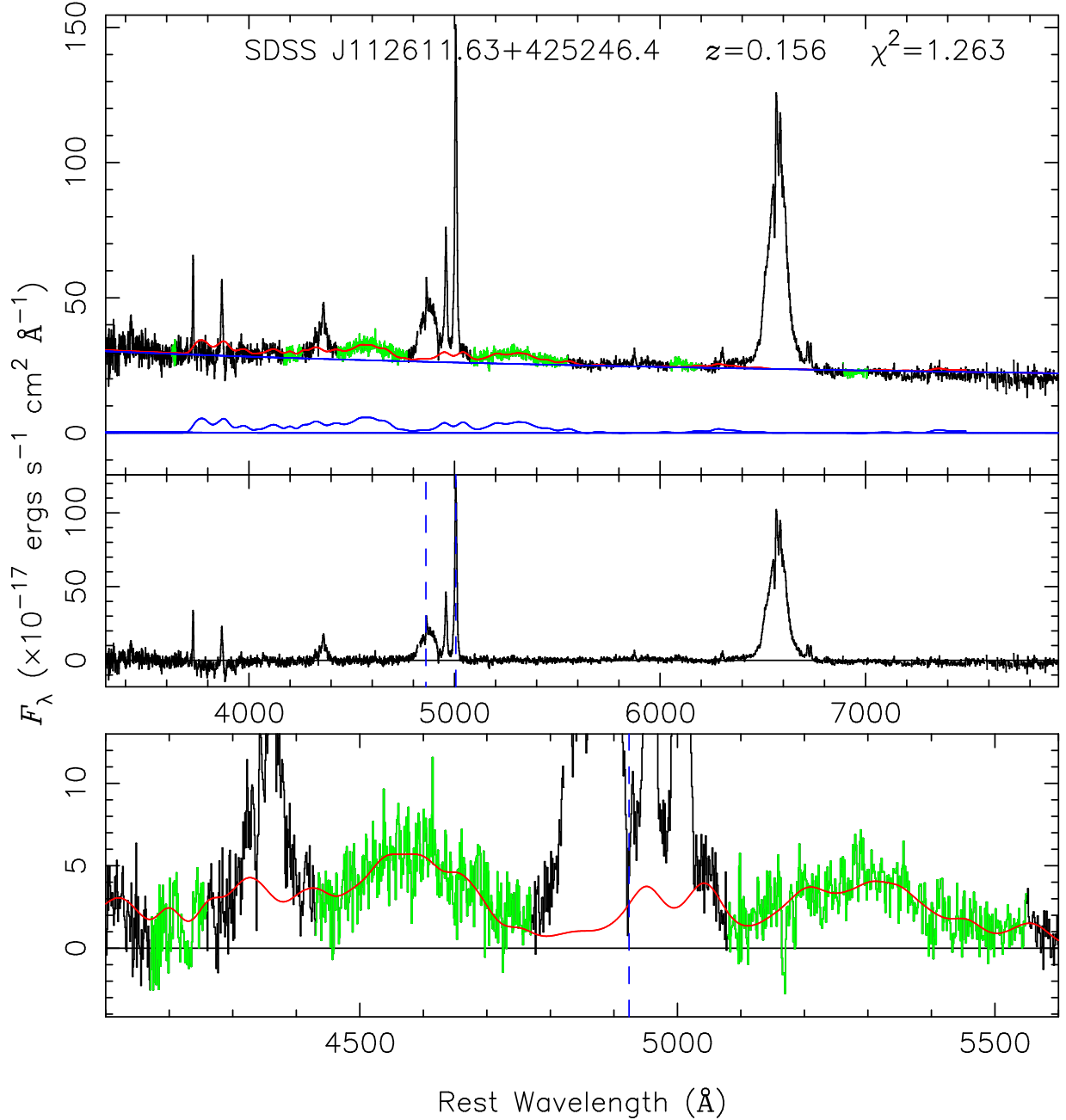}
  \caption{ Same as Fig. \ref{fig-conti1}, but for 
  SDSS J112611.63+425246.4, which has broad Fe lines.
  Note the large departure of the modeled \feii\ $\lambda$4924 line from its
  rest-frame wavelength (blue dashed line in the bottom panel). \vfe\ is
  1691$\pm$119 \kms.}
  \label{fig-conti2}
\end{figure*}

Two examples of continuum decomposition are shown in Figures \ref{fig-conti1}
and \ref{fig-conti2}.
The top panel shows the corrected spectrum.  The spectrum in the fitting
window is plotted in green. Each component is plotted in blue and the summed
continuum in red.  The middle panel shows the residual spectrum, which is
the pure emission-line spectrum for the next step (\S \ref{linefit}). The
model fits the spectrum very well except in the region $\sim$ 3100--3700
\ang, where no Fe template is available.  In fact, the total flux of the
residual spectrum in this region strongly correlates with the Fe flux $F_{\rm
Fe}$.  This is consistent with the ``small blue bump'' being produced by Fe 
lines and the Balmer continuum \citep{wil85}. We subtract the power law and the
Balmer continuum from the corrected spectrum and show the enlarged resultant
spectrum (Fe-only spectrum) in the wavelength range 4100--5600 \ang\ in the
bottom panel. The Fe model is plotted in red.  

From the first example on SDSS J115507.61+520129.6, which has 
narrow Fe lines, the Fe model not only agrees with the Fe-only spectrum in our 
fitting windows (in green), but it also fits the two strong \feii\ lines at 
4924 and 5018 \ang\ very well, even though these two lines are not in the
fitting windows. Two blue dashed lines in the middle panel mark the positions
of \hb\ $\lambda$4861 and \oiii\ $\lambda$5007. The blue dashed line in the 
bottom panel marks the position of \feii\ $\lambda$4924 line with zero 
velocity shift. The shift of Fe emission in this source can be seen clearly 
from the position of the \feii\ $\lambda$4924 line; the velocity shift, as 
measured from the model and with respect to \oiii,  is 459$\pm$16 \kms.  For 
the second source, SDSS J112611.63+425246.4, \feii\ $\lambda$4924 cannot be 
distinguished from \oiii\ $\lambda$4959 but the shift can be seen from the 
\feii\ model. In this case, it is 1691$\pm$119 \kms.

\subsection{Emission-line Fitting}
\label{linefit}

\begin{figure*}
  \centering
  \includegraphics[angle=-90,width=0.45\textwidth]{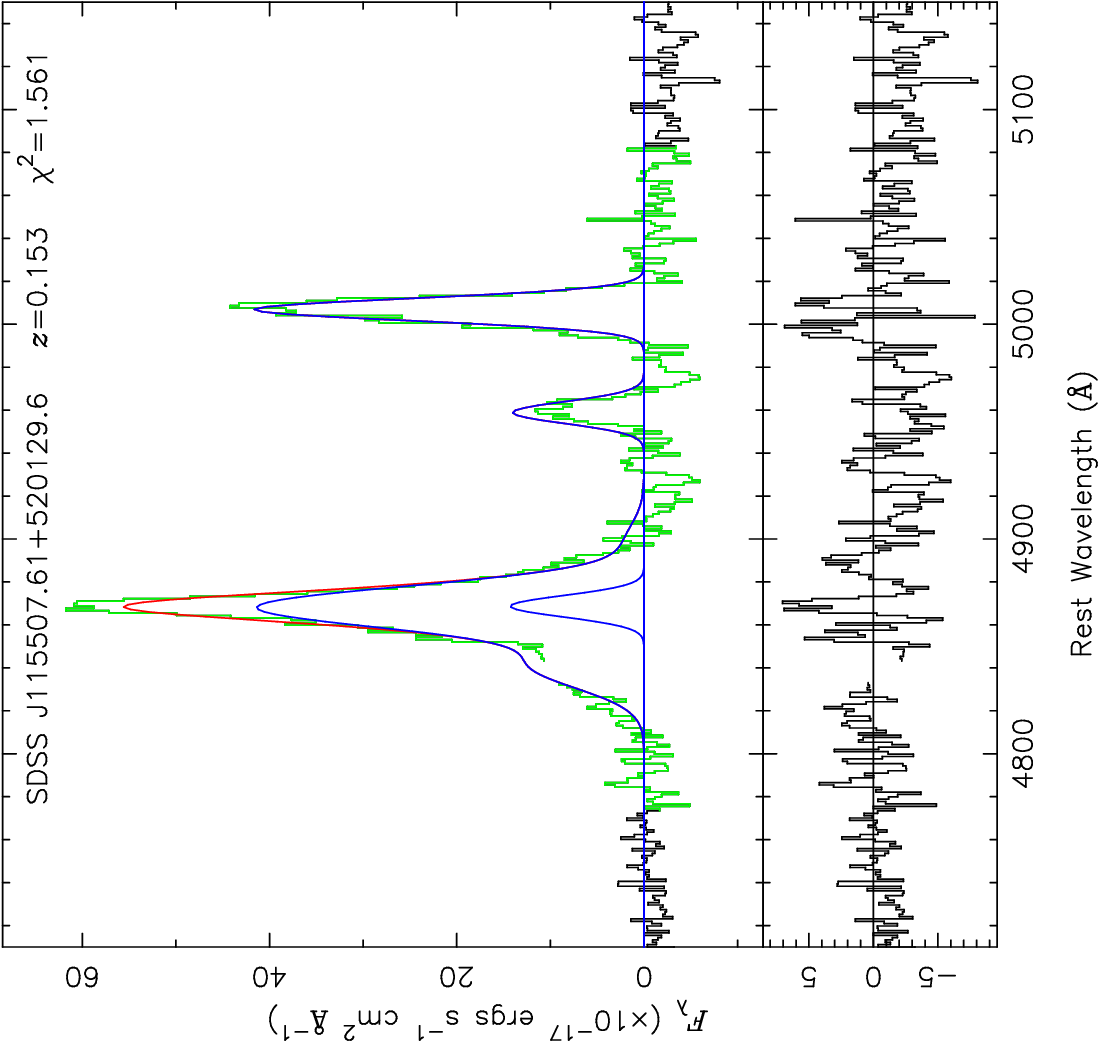}
  \hspace{0.4cm}\includegraphics[angle=-90,width=0.45\textwidth]{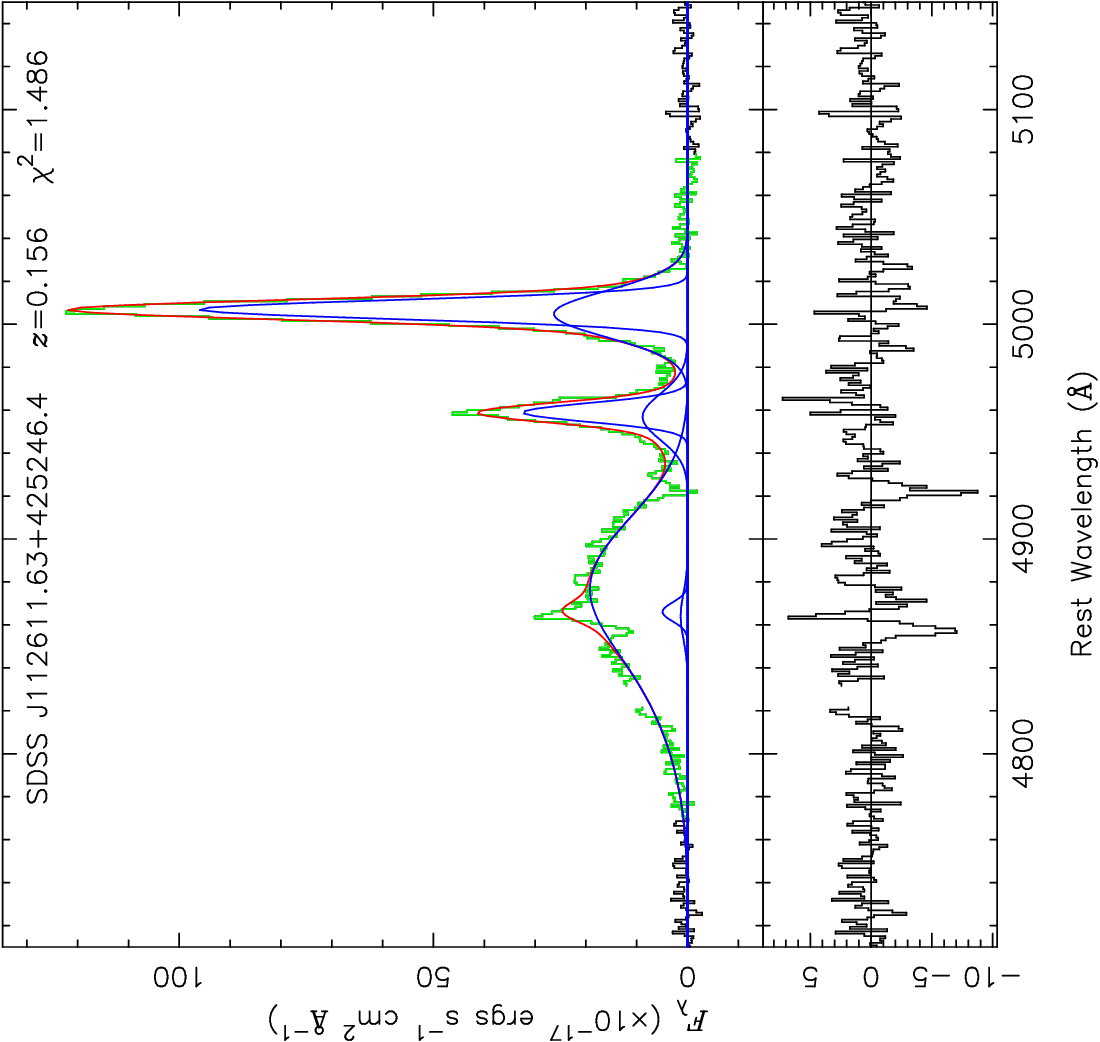}
  \caption{ Fitting of the emission lines for the two quasars in 
  Figs. \ref{fig-conti1} and \ref{fig-conti2}. The top panel shows the pure 
  emission-line spectrum. Each component is in blue, 
  and the sum of them is in red. The bottom panel shows the residuals.}
  \label{fig-hbo3}
\end{figure*}

After subtracting the continuum, we measure the \hb\ and \oiii\
emission lines from the pure emission-line spectrum. We use multiple
components to fit the emission lines over the wavelength range 4770--5080 \ang.
The narrow \hb\ component, \oiii\ $\lambda$4959, and \oiii\ $\lambda$5007 are
modeled using three Gaussian. The \oiii\
$\lambda\lambda$4959, 5007 lines are forced to have the same FWHM and no 
relative wavelength shift, and their intensity ratio fixed to the theoretical 
value of 3.0. The narrow \hb\ component (\hbn) is forced to have the same FWHM 
as \oiii\ $\lambda$5007, a shift of up to 600 \kms\ relative to \oiii\ 
$\lambda$5007, and an intensity constrained to lie between 1/20 and 1/3 of that
of \oiii\ $\lambda$5007 \citep[e.g.][]{veilleux87,mcgill08}.  If 
necessary, we add another two Gaussian components for 
\oiii\ $\lambda\lambda$4959, 5007  to match their wings, and a
corresponding Gaussian is added to \hbn\ to ensure that \hbn\ and \oiii\
have the same profile. Following \citet{salv07} and \citet{mcgill08}, the broad \hb\
component is modeled using a Gauss-Hermite function \citep{vandermarel93},
whose best fit yields the FWHM, peak velocity shift, and the square root of
the second moment ($\sigma_{{\rm H}\beta}$). As illustration, Figure 
\ref{fig-hbo3} shows the emission-line fitting for the two sources in Figures
\ref{fig-conti1} and \ref{fig-conti2}.
The top panel shows the pure emission-line spectrum. The multiple  
components are in blue and the sum of them is in red; the bottom panel is the 
residuals.  

We also measure the \oii\ and \mgii\ emission lines when they are available.
Since the continuum around \oii\ is not well determined (we make two
assumptions in the model for the higher-order Balmer lines; see the last 
paragraph in \S \ref{balconti}), we simply fit \oii\ above a locally defined
continuum with a single Gaussian \citep{greene05a}. As many sources have weak
\oii\ emission, we adopt as detection criterion that the line must have an
amplitude larger than 3 times the standard deviation of the local continuum.
The \mgii\ $\lambda\lambda$2796, 2803 doublet is fitted using two
Gauss-Hermite functions; they have the same parameters except that the
intensity ratio between them is fixed to 2 \citep{bal96}. The FWHMs and
velocity shifts of \mgii\ are calculated from a single Gauss-Hermite function.

\subsection{Tests of the Continuum Decomposition}
\label{mcsim}
The template-fitting method for measuring Fe emission and the 
Levenberg-Marquardt algorithm for nonlinear fitting are widely used as almost 
``standard'' approaches. We add a new parameter \vvfe\ and set \fwhmfe\ free 
in our fitting. Since the S/N of the majority of the sources in our sample are 
low (about 50\% have S/N $<$ 15), it is necessary to test the reliability of
the Fe emission measurement.  We perform a suite of simulations similar to those
done in \citet{greene06b}.  Using a Monte Carlo method to generate artificial 
spectra, we measure the Fe emission of these spectra using the same method as
that used for the observed spectra.  Differences between input and output 
parameters can then be compared to evaluate potential systematic errors and 
biases.

We build the simulated continuum spectrum as a linear combination of a
single power law and Fe emission expressed by Eq. (\ref{equ-fe}). 
We generate a realistic noise pattern for the spectra using a real error 
array taken from SDSS observations, scaling it by a multiplicative factor 
to match the desired S/N of the simulation.  For each pixel of the
simulated spectrum, a Gaussian random deviation is added. Bad pixels have
large deviates statistically.  We use a mask array from an actual FITS file 
to locate the masked pixels; some of the pixels that have large errors are 
masked by the SDSS pipeline, but not all.  This procedure ensures that the 
simulated spectra have a realistic noise level and noise pattern. 

There are three main factors that can affect the measurements of \vvfe\ and
\fwhmfe: (1) the S/N of the spectrum; (2) the strength (EW) of the iron 
emission; and (3) the width (FWHM) of the iron lines.  Our simulations
demonstrate that the input value of \vvfe\ in the simulated spectrum 
has a very minimal effect on the systematic bias of the measured output 
values of \vvfe\ or \fwhmfe, and so we neglect it from further consideration.  
We calculate 
${\rm EW_{Fe}}=F({\rm Fe}~\textsc{ii}~\lambda4570)/F_{5100}$ where, in the 
present paper, $F({\rm Fe}~\textsc{ii}~\lambda4570)$ is the flux of the \feii\ 
emission between 4434 and 4684 \AA.  Most of the quasars in the sample have a 
\ewfe\ between 15 and 75 \ang\ and \fwhmfe\ between 1000 and 5000 \kms. 
Accordingly, we test four values of \ewfe\ (15, 25, 50, 75 \ang) and vary 
\fwhmfe\ from 1000 to 5000 \kms, in steps of 500 \kms. We set S/N = 10, which 
is the lower limit of the S/N in the sample.  \vvfe\ is fixed at zero so that 
we can examine whether the measured shifts are real or spurious.

\begin{figure*}
  \centering
  \includegraphics[angle=-90,width=0.45\textwidth]{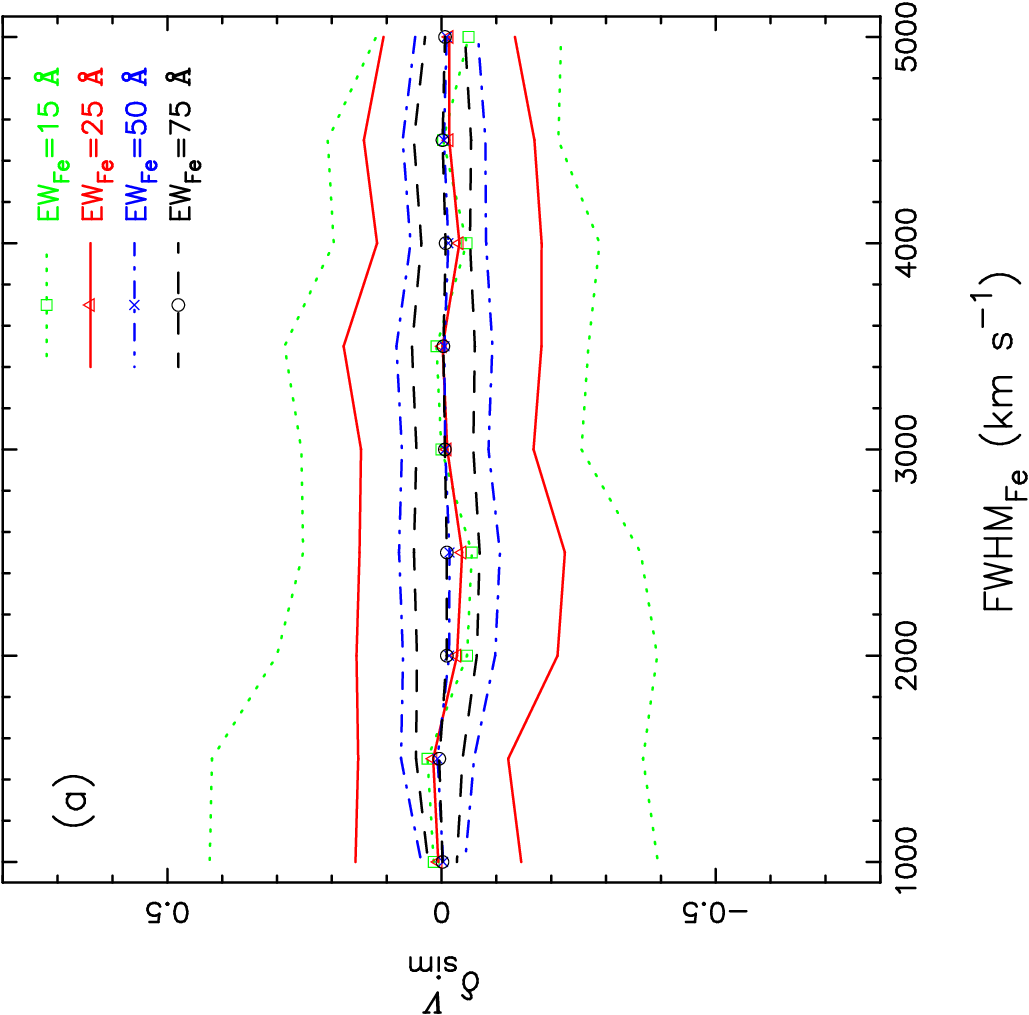}%
  \hspace{0.4cm}\includegraphics[angle=-90,width=0.45\textwidth]{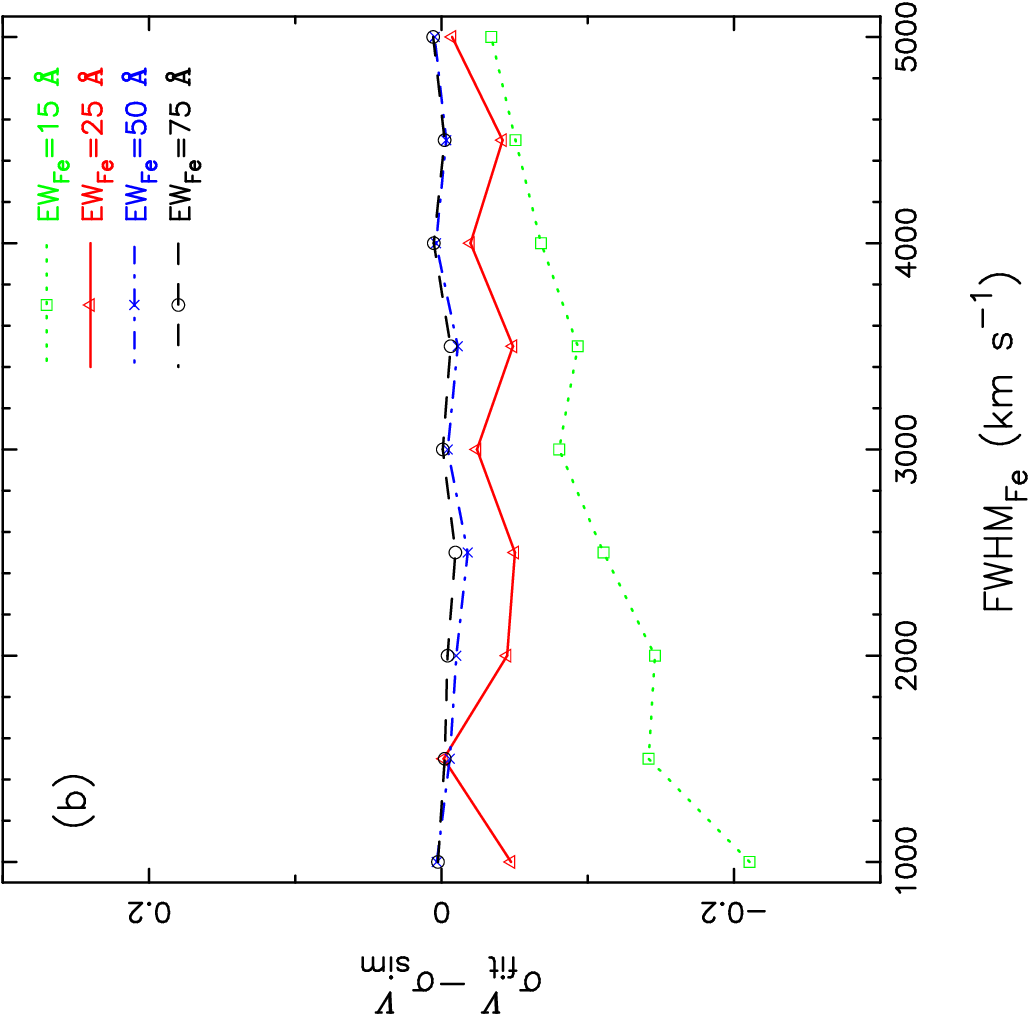}
  
  \vspace{0.5cm}
  \includegraphics[angle=-90,width=0.45\textwidth]{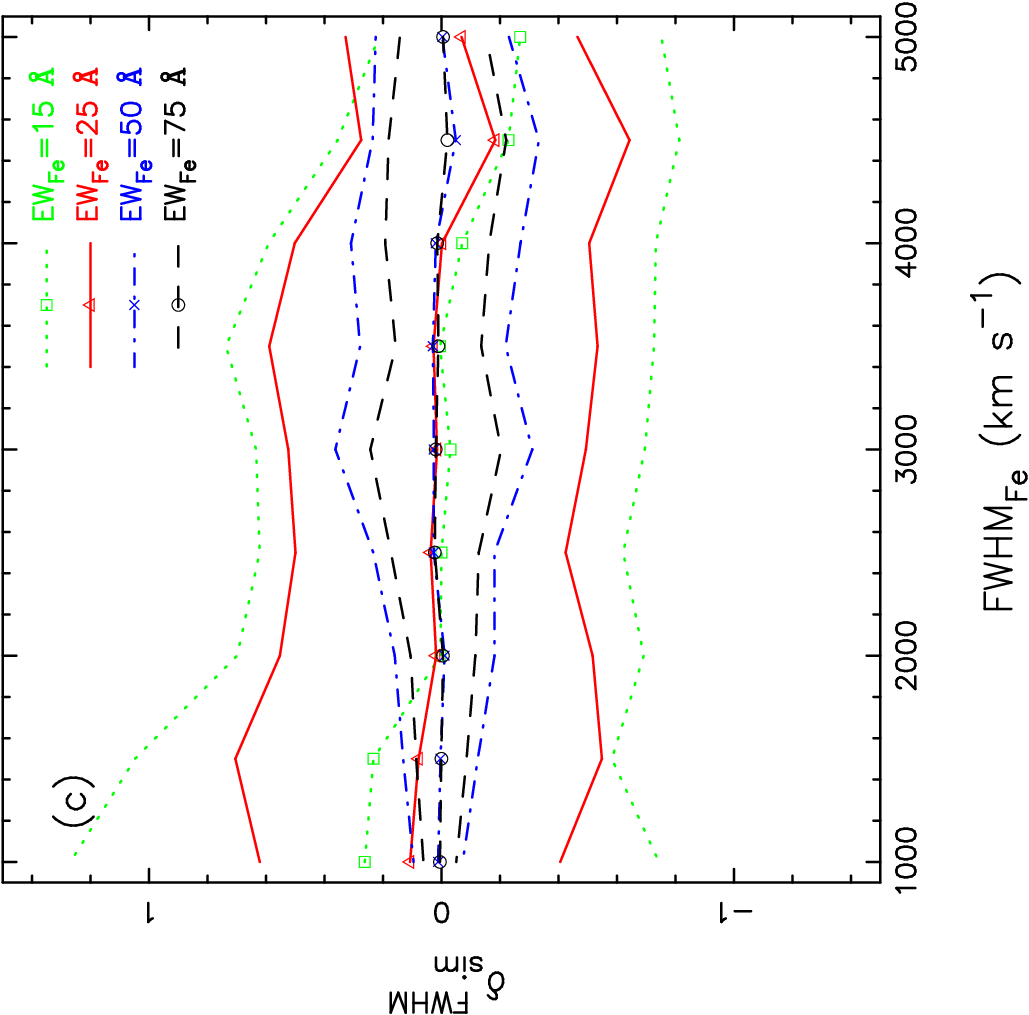}%
  \hspace{0.4cm}\includegraphics[angle=-90,width=0.45\textwidth]{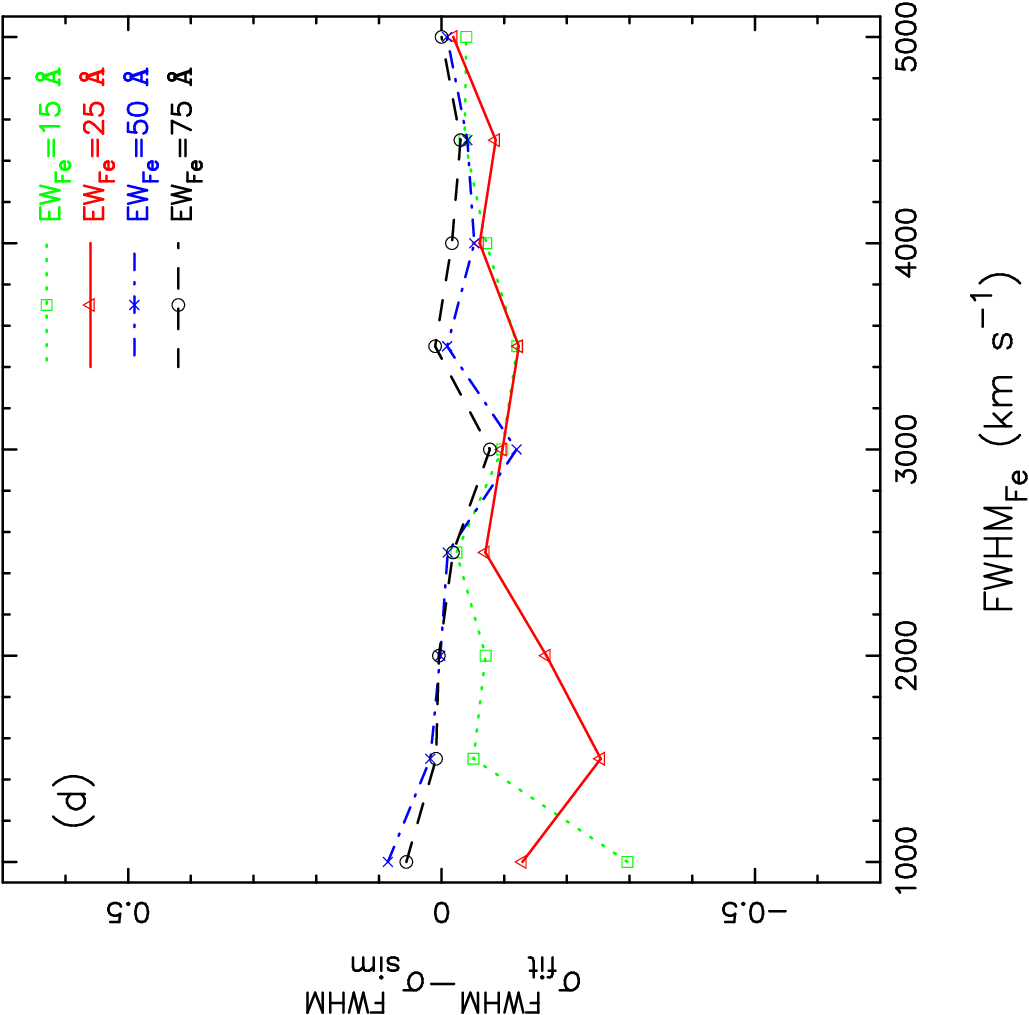}
  \caption{ ({\it a}) Input-output simulations of \vfe\
  measurements. We define \dsv\ $=(V_{\rm in}-V_{\rm out})/{\rm FWHM_{in}}$. 
  For each \ewfe, the middle line represents the mean \dsv, and the other two 
  lines above and below are one standard deviation above and below the mean. The
  S/N in the simulation is 10, which is the lower limit of our sample, so the
  uncertainties shown here should be considered upper limits.  ({\it b}) Test 
  of the error bars given by the code.  \sfv\ is the average value of the 
  errors given by the code in 100 trials, divided by ${\rm FWHM_{in}}$, and 
  \ssv\ is the standard deviation of \dsv.  ({\it c}) Input-output simulations 
  of \fwhmfe\ measurements. We define \dsw\ 
  $={\rm (FWHM_{out}-FWHM_{in})/FWHM_{in}}$. ({\it d}) Test of the error bars 
  by the code. \sfw\ is the average value of the relative errors of \fwhmfe\ 
  given by the code in 100 trials, and \ssw\ is the standard deviation of 
  \dsw. See text for details of the simulation.}
  \label{fig-mc}
\end{figure*}

For each pair of values for \ewfe\ and \fwhmfe, we generate 100 spectra. We 
fit the continuum and then calculate the quantity
\begin{equation}
  \delta_{\rm sim}^V=\frac{V_{\rm in}-V_{\rm out}}{{\rm FWHM}_{\rm in}},
\end{equation}
where $V_{\rm in}$ and $V_{\rm out}$ are input and output values of $V_{\rm
Fe}$, respectively, and ${\rm FWHM_{in}}$ is the input \fwhmfe. 
Figure \ref{fig-mc}{\it a}\ shows \dsv\ as a function of \fwhmfe.  
The results are plotted using different colors and line styles for different 
values of \ewfe. For each \ewfe, there are three lines: the middle line 
represents the mean \dsv, and the other two lines above and below are one 
standard deviations (\ssv)
above and below the mean. From the diagram, we can see that the average value
of \dsv\ is always close to 0.   This demonstrates that 
the measured value of \vvfe\ shows no systematic redshift or blueshift with
increasing \fwhmfe\ and \ewfe. 
Obviously, \ssv\ increases with decreasing \ewfe; measuring \feii\ emission 
is very uncertain when \ewfe\ is small.  Based on the results of these 
tests, we decided to exclude from the sample sources with \ewfe\ $<$ 25 \ang.
It should be noted that the S/N in this particular simulation is
set at the lower limit of our sample, so the uncertainties shown in Figure
\ref{fig-mc} should be considered upper limits.  Figure \ref{fig-mc}{\it b}\ 
shows the difference between the error given by our code and that given by the 
simulation. \sfv\ is the average value of the
errors given by the code in the 100 trials divided by ${\rm FWHM_{in}}$. 

Figures \ref{fig-mc}{\it c}\ and \ref{fig-mc}{\it d}\ show the systematic error
in the measurement of \fwhmfe. We define
\begin{equation}
\delta_{\rm sim}^{\rm FWHM}={\rm \frac{FWHM_{out}}{FWHM_{in}}-1}.
\end{equation}
The standard deviation of \dsw\ is denoted by \ssw, and \sfw\ is the average 
value of the relative errors of \fwhmfe\ (defined as the error of the \fwhmfe\ 
divided by ${\rm FWHM_{in}}$) given by the code.

From these simulations, we conclude: (1) the mean values of \dsv\ and \dsw\ 
cluster around 0 and exhibit no trend as a function of \fwhmfe\ or \ewfe; (2) 
the error given by our code is consistent with that given by the simulations 
(except perhaps for the \ewfe\ $=$ 15 \ang\ bin). 
These results show that our measurements are reliable and robust.

\subsection{Comparison with Conventional Fe Template-fitting Methods}
\label{compare}
There are two conventional methods for Fe fitting.  Both have been widely 
used and effective.   The first (model 1) assumes that the Fe lines have no 
velocity shift but that their width can be different with that of \hbb\ 
\citep[e.g.,][]{bg92,marzi96,mclure02,die03,greene05b,kim06,woo06}.
Alternatively (model 2), one assumes that the Fe lines have no shift and that 
they have the same width as \hbb\ \citep[e.g.,][]{netzer07,salv07}.  Two 
obvious questions arise.  Does our continuum model (Eq. (\ref{equ-conti})) 
significantly improve the fit? And second, how does our continuum 
decomposition affect the emission-line measurements (e.g., \hb\ and \oiii) and 
the physical parameters subsequently derived from them (e.g., central BH mass 
and the Eddington ratio)?

We compare the results derived from our approach with the two standard 
methods described above.  Following \citet[Chapter 12.1]{lupton93}, we use the
F-test\footnote{Strictly speaking, this test is valid only for models that 
use linear fitting. But it has been empirically used for nonlinear models and 
seems to be effective (e.g., see \citealt{hao05}).} to calculate how 
significantly our model improves the fit for each source. Comparing with model 
1, 71\% of the sources are better fit by our model at a significance of 
$>$95.45\%, and 58\% of the sources are better fit at a significance level of 
$>$99.73\%.  With respect to model 2, the corresponding improvement can be 
seen in 89\% of the sources at a significance of $>$95.45\% and in 80\% of the 
sources at a significance of $>$99.73\%.  On average, our model decreases the 
reduced \ks\ by 0.039 and 0.094 compared  with models 1 and 2, respectively. 
We conclude that our approach of allowing \vvfe\ and \fwhmfe\ to be free 
parameters significantly improves the fit in most objects.

Next, we evaluate the actual impact that the different methods have on measured 
and derived physical quantities.  We measure the \hbb\ and \oiii\ emission 
lines, derive \mbh\ and \er\ (see \S \ref{resulter}) for each model, and then 
calculate the relative differences of the parameters between our model and the
two fiducial standard models.  As summarized in Table \ref{tab-diff}, the 
differences in line luminosities and line widths for \hbb\ and \oiii, \mbh,
and \er\ are all less than 5\%, while changes in velocity shifts are also no 
more than 50 \kms.  The only exception is for sources with very weak \oiii\ 
lines.  In this regime, the measurement of \oiii\ can be strongly affected by 
\feii\ $\lambda$4924 and \feii\ $\lambda$5018, and the effect on 
$L_{\textsc{[O iii]}}$ and FWHM$_{\textsc{[O iii]}}$ in Table \ref{tab-diff} is 
large (a few tens of percent).  This exercise demonstrates that, for most 
applications, the choice of method for Fe template fitting is in practice 
unimportant---unless the main scientific objective is to actually study the 
\feii\ emission itself. 

\begin{deluxetable*}{ccccccccc}
  \tablewidth{0pt}
  \tablecolumns{9}
  \tablecaption{Effect of Fe Template-fitting Method on Other Parameters
  \label{tab-diff}}
  \tablehead{
  \colhead{Model} & \colhead{$L({\rm H}\beta_{\rm BC})$} &
  \colhead{\fwhmhbb} & \colhead{\vvhbb} & \colhead{$L_{\textsc{[O iii]}}$} &
  \colhead{FWHM$_{\textsc{[O iii]}}$} & \colhead{$V_{\textsc{[O iii]}}$} & 
  \colhead{\mbh} & \colhead{\er} \\
  \colhead{(1)} & \colhead{(2)} & \colhead{(3)} & \colhead{(4)} &
  \colhead{(5)} & \colhead{(6)} & \colhead{(7)} & \colhead{(8)} &
  \colhead{(9)}
  }
  \startdata
  \scriptsize
  fix \vvfe, free \fwhmfe\     & $-$1.75\%(2.99\%) & $-$1.74\%(3.54\%) & 
  $-$27.2(49.2) & $-$2.15\%(5.01\%) & $-$1.02\%(9.51\%) & 2.76(35.6) &
  $-$0.07\%(0.49\%) & 2.23\%(11.2\%) \\
  \scriptsize
  fix both \vvfe\ and \fwhmfe\ & $-$1.46\%(3.14\%) & $-$0.39\%(4.10\%) &
  $-$23.2(47.9) & $-$0.48\%(5.58\%) & $-$0.31\%(10.22\%) & 2.95(32.0) &
  $-$0.01\%(0.46\%) & 0.63\%(10.5\%) 
  \enddata

  \tablecomments{Changes in emission-line parameters and derived physical
  parameters due to using different Fe template-fitting models. Col. (2):
  luminosity of \hbb. Col. (3): FWHM of \hbb. Col. (4): \hbb\ velocity shift.
  Col. (5): luminosity of \oiii. Col. (6): FWHM of \oiii. Col. (7): \oiii\
  velocity shift. Col. (8): mass of the central BH. Col. (9): Eddington ratio 
  of the central BH (see \S \ref{resulter} for details of how
  to derive \mbh\ and \er). Each column shows the relative changes in
  percentage, except for Col. (4) and (7), which show the difference in 
  absolute velocity shift in \kms. The number in parenthesis is the standard 
  deviation.}
\end{deluxetable*}

\subsection{Redshifts}
\label{redshift}
\begin{figure}
  \centering
  \includegraphics[angle=-90,width=0.45\textwidth]{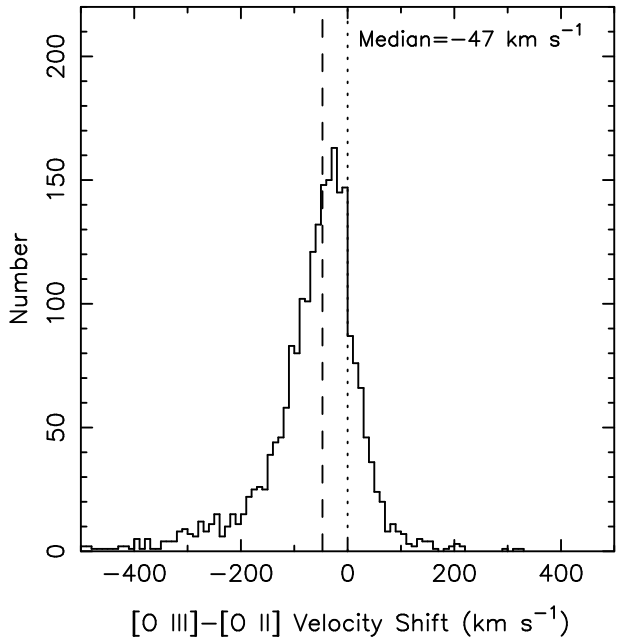}
  \caption{ Distribution of \oiii\ velocity shifts with respect to
  \oii. Positive velocity indicates a redshift. The dotted line marks the
  position of zero velocity shift, while the dashed line marks the median of the
  distribution.}
  \label{fig-o2shift}
\end{figure}

The narrow emission lines are commonly used to obtain the systemic redshift. 
The \oiii\ $\lambda$5007 line is the strongest narrow line for most quasars, 
so it is most often used. However, \citet{boroson05} showed that 
\oiii\ can be blueshifted with respect to the low-ionization forbidden lines, 
which provide a better rest-frame. In our sample, 2265 sources ($\sim 50$\%) 
have detectable \oii\ $\lambda$3727 emission.  Figure \ref{fig-o2shift} shows 
the distribution of velocity shifts between \oiii\ and \oii; positive velocity 
indicates a redward shift. Consistent with the results of 
\citet{boroson05}, the majority of the sources have blueshifted \oiii; the 
median blueshift is $-47$ \kms.  Because the \oiii\ blueshifts are much 
smaller than the velocity shifts seen in \feii\ and in other broad lines (see 
\S \ref{resultshift}) and only half of our sample is detected in \oii, we 
still use \oiii\ to define the rest-frame.  Thus, we define the \feii\ 
velocity shift by \vfe=\vvfe$-V_{\textsc{[O iii]}}$, where $V_{\textsc{[O iii]}}$ is the 
velocity shift of the core component of the \oiii\ $\lambda5007$ line measured 
in \S \ref{linefit}.  

\subsection{Errors}
\label{error}

\begin{figure*}
  \centering
  \includegraphics[angle=-90,width=0.9\textwidth]{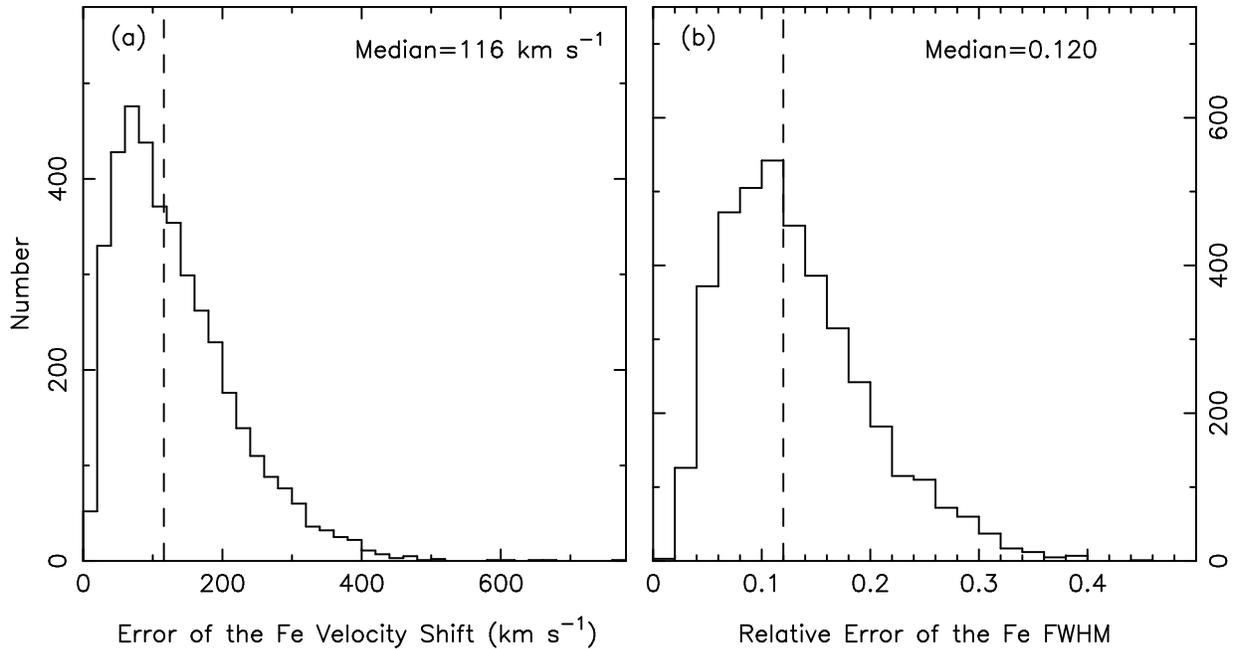}
  \caption{ Distribution of ({\em a}) error in the \feii\
  velocity shift and ({\em b}) relative error in \feii\ FWHM. The dashed
  line marks the median of the distribution.}
  \label{fig-err}
\end{figure*}

We calculate the error of \vfe\ from the fitting of the continuum and
the fitting of the \oiii\ line. Figure \ref{fig-err}{\em a} shows the
distribution of the \vfe\ errors. Most velocity shifts have an error $< 200$ 
\kms, and the median of the distribution is 116 \kms. We also plot the
relative error on \fwhmfe\ in Figure \ref{fig-err}{\em b}.  The typical value 
is about 10\% to 20\%, and the median is 12.0\%. 

\section{Results}
\label{result}

The full catalog of the measurements used in the analysis below is available
electronically. Table \ref{tab-format} describes the contents and the
formats of each column. All the emission line velocity shifts are with
respect to \oiii\ $\lambda$5007 and a positive value indicates a redward
shift. $-9999$ in the columns of \mgii\ (Col. 21---26) and \oii\
(Col. 35---40) measurements indicates the line lies out of the SDSS spectral
coverage or too weak to be detected (see \S \ref{linefit} for details). The
details of how to deriving the radio and X-ray properties (Col. 19 \& 20),
and the \mbh\ and Eddington ratio \er\ (Col. 42 \& 43), are described in \S
\ref{resultother} and \S \ref{resulter} respectively.

\begin{deluxetable*}{lcccl}
  \tablewidth{0pt}
  \tablecolumns{3}
  \tablecaption{Spectrophotometric Measurements Table Format
  \label{tab-format}}
  \tablehead{
  \colhead{Column} & \colhead{Format} & \colhead{Units} & \colhead{Label}
  &\colhead{Description} 
  }
  \startdata
  1  & a18   & \nodata & SDSS Name & SDSS DR5 Object Designation hhmmss.ss$+$ddmmss.s (J2000.0) \\
  2  & f8.4  & \nodata & $z$       & Redshift determined using the peak of the \oiii\ $\lambda$5007 \\
  3  & e12.4 & \ergs   & $L({\rm Fe} \textsc{ ii } \lambda4570)$  & Luminosity of the \feii\ emission between 4434 and 4686 \ang \\
  4  & e12.4 &         &           & Error in $L({\rm Fe} \textsc{ ii } \lambda4570)$ \\
  5  & f8.1  & \kms    & \fwhmfe   & \feii\ FWHM \\
  6  & f8.1  &         &           & Error in \fwhmfe \\
  7  & f8.1  & \kms    & \vfe      & \feii\ velocity shift; all the emission line velocity shifts are \\
                             & & & & ~~~~with respect to \oiii\ $\lambda$5007, and a positive value indicates a redward shift.\\
  8  & f8.1  &         &           & Error in \vfe \\
  9  & e12.4 & \ergs   & $L({\rm H}\beta_{\rm BC})$ & \hbb\ luminosity \\
  10 & e12.4 &         &           & Error in $L({\rm H}\beta_{\rm BC})$ \\
  11 & f8.1  & \kms    & \fwhmhbb  & \hbb\ FWHM \\
  12 & f8.1  &         &           & Error in \fwhmhbb \\
  13 & f8.1  & \kms    & \vhbb     & \hbb\ velocity shift \\
  14 & f8.1  &         &           & Error in \vhbb \\
  15 & e12.4 & \ergsa  & $L_{5100}$ & Specific continuum luminosity at 5100 \ang \\
  16 & e12.4 &         &           & Error in $L_{5100}$ \\
  17 & f7.3  & \nodata & $\alpha$  & Power law spectral index of the continuum \\
  18 & f7.3  & \nodata &           & Error in $\alpha$ \\
  19 & a2e10.3 & \ergsh & $L_{\rm 6 cm}$ & Specific luminosity at 6 cm derived from the Peak flux density measured in FIRST, \\
                             & & & & ~~~~assuming a radio spectral index = $-$0.5; the upper limits are derived from \\
                             & & & & ~~~~the FIRST flux limits; $-$1.000 indicates not in FIRST survey area. \\ 
  20 & e11.3 & \ergsh  & $L_{\rm 2 keV}$ & Specific luminosity at 2 kev derived from {\it ROSAT}\ PSPC count rate;\\
                             & & & & ~~~~0.000 means no detection. \\
  21 & e12.4 & \ergs   & $L_{{\rm Mg} \textsc{ ii}}$ & \mgii\ luminosity; $-$9999 means no detection. \\
  22 & e12.4 &         &           & Error in $L_{{\rm Mg} \textsc{ ii}}$ \\
  23 & f8.1  & \kms    & ${\rm FWHM}_{{\rm Mg} \textsc{ ii}}$ & \mgii\ FWHM \\
  24 & f8.1  &         &           & Error in ${\rm FWHM}_{{\rm Mg} \textsc{ ii}}$ \\
  25 & f8.1  & \kms    & $v_{{\rm Mg} \textsc{ ii}}$  & \mgii\ velocity shift \\
  26 & f8.1  &         &           & Error in $v_{{\rm Mg} \textsc{ ii}}$ \\
  27 & e12.4 & \ergs   & $L_{\textsc{[O iii]}}$ & \oiii\ $\lambda$5007 luminosity \\
  28 & e12.4 &         &           & Error in $L_{\textsc{[O iii]}}$ \\
  29 & f8.1  & \kms    & ${\rm FWHM}_{\textsc{[O iii]}}$ & \oiii\ $\lambda$5007 FWHM \\
  30 & f8.1  &         &           & Error in ${\rm FWHM}_{\textsc{[O iii]}}$ \\
  31 & e12.4 & \nodata & $L({\rm H}\beta_{\rm NC})/L_{\textsc{[O iii]}}$ & Ratio of \hbn\ to \oiii\ $\lambda$5007 \\
  32 & e12.4 & \nodata &           & Error in $L({\rm H}\beta_{\rm NC})/L_{\textsc{[O iii]}}$ \\
  33 & f8.1  & \kms    & \vhbn     & \hbn\ velocity shift \\
  34 & f8.1  &         &           & Error in \vhbn \\
  35 & e12.4 & \ergs   & $L_{\textsc{[O ii]}}$ & \oii\ luminosity; $-$9999 means no detection. \\
  36 & e12.4 &         &           & Error in $L_{\textsc{[O ii]}}$ \\
  37 & f8.1  & \kms    & ${\rm FWHM}_{\textsc{[O ii]}}$  & \oii\ FWHM \\
  38 & f8.1  &         &           & Error in ${\rm FWHM}_{\textsc{[O ii]}}$ \\
  39 & f8.1  & \kms    & $v_{\textsc{[O ii]}}$ & \oii\ velocity shift \\
  40 & f8.1  &         &           & Error in $v_{\textsc{[O ii]}}$ \\
  41 & f8.1  & \kms    & $\sigma_{{\rm H}\beta}$ & Square root of the second moment of \hbb \\
  42 & e11.3 & $M_\odot$ & \mbh    & Mass of the central Black Hole \\
  43 & e11.3 & \nodata & \er       & Ratio of bolometric luminosity to Eddtington luminosity \\
  44 & e12.4 & \nodata & \rfe      & Ratio of \feii\ to \hbb\ \\
  \enddata
\end{deluxetable*}

\subsection{Fe Emission Shifts}
\label{resultshift}
\begin{figure*}
  \centering
  \includegraphics[angle=-90,width=0.9\textwidth]{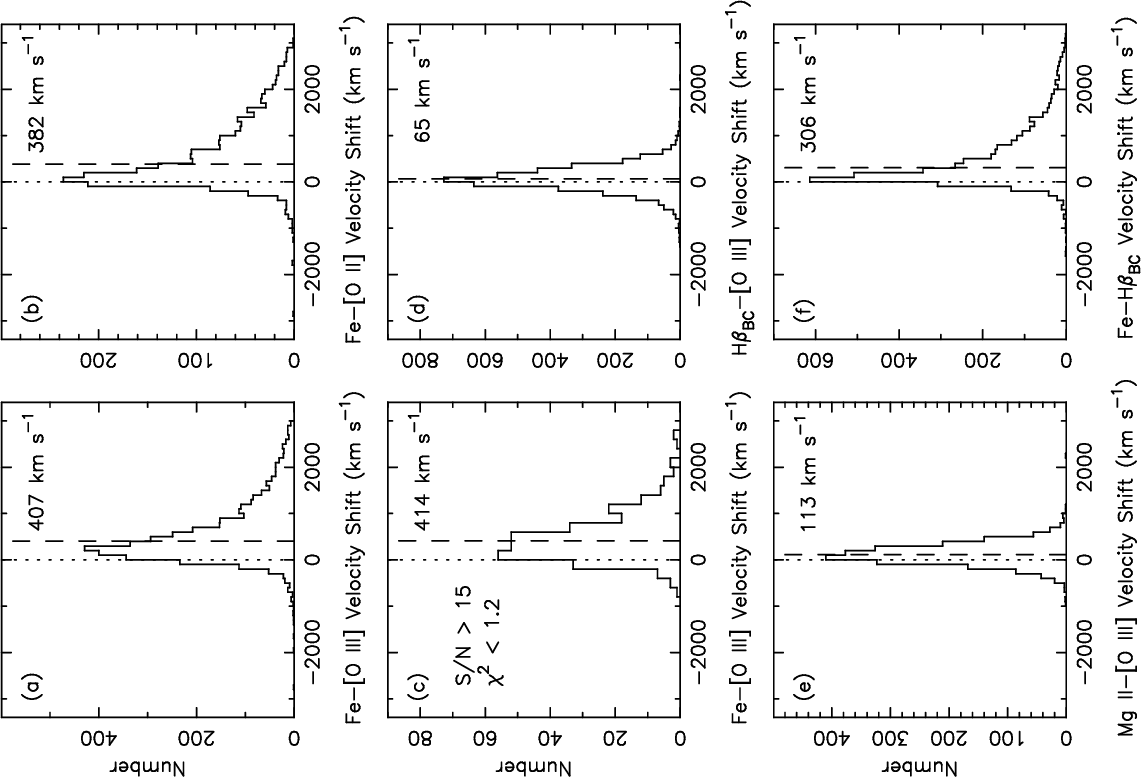}
  \caption{ ({\it a}) Distribution of \feii\ velocity shifts with
  respect to \oiii.  ({\it b}) \feii\ velocity shifts with respect to \oii.  
  Panel ({\it c}) shows the distributions of \vfe\ in the subsample that has 
  stricter criteria S/N $>$ 15 and \ks\ $<$ 1.2.  ({\it d}) \hbb\ velocity 
  shifts with respect to \oiii.  ({\it e}) \mgii\ velocity shifts with respect 
  to \oiii.  ({\it f}) Distribution of \feii\ velocity shifts with respect to 
  \hbb.  Positive velocity indicates a redshift.  The number in each panel is 
  the median value of the respective distribution, which is also marked by 
  the dashed line.  The dotted line marks the position of zero velocity shift.
}
  \label{fig-shift}
\end{figure*}

From the distribution of \vfe\ (Fig. \ref{fig-shift}{\it a}), we can clearly 
see that most quasars exhibit \feii\ emission that is {\it redshifted}\ 
with respect to the systemic velocity of the narrow-line region (defined 
by \oiii).  Only 481 out of 4037 quasars have blueshifts.  The median shift is
\vfe\ = $+$407 \kms\ (the vertical dashed line in the figure), with a maximum 
value of \vfe\ $\approx$ 2000 \kms. Considering that the typical error on 
\vfe\ is only $<$200 \kms\ (\S \ref{error}), the vast majority of the values
in the redshifted tail of the \vfe\ distribution must be real.  For comparison,
we also calculate the distribution of \feii\ velocity shift with respect to 
\oii\ (Fig. \ref{fig-shift}{\it b}).  The results obtained by using \oii\ 
as reference instead of \oiii\ are very similar and lead to the same 
conclusion.

To rule out the possibility that the excess \feii\ redshifts arise from 
artifacts due to poor data quality or fitting errors, we examined a subset of 
data using the much stricter selection criteria that S/N $>$ 15 and \ks $<$ 
1.2.  These 309 quasars have the best data quality and the most reliable
continuum fitting.  The shape of the distribution of \vfe\ this 
subsample (Fig. \ref{fig-shift}{\it c}) is quite similar to that of our whole 
sample.  The median velocity shift is 414 \kms.   We conclude on the basis of 
this test, as well as the Monte Carlo simulations described in \S \ref{mcsim},
that the \feii\ redshifts are robust and reliable.

Finally, we illustrate that the velocity shifts found for \feii\ really do 
imply radial motions of the \feii-emitting region, and not others.  Figures  
\ref{fig-shift}{\it d}\ and \ref{fig-shift}{\it e}\ show the velocity shifts 
for \hbb\ [\vhbb] and \mgii.  The distribution of \vhbb\ is almost symmetrical 
around 0, with a median value of only 65 \kms\ and a maximum value of about 
$\pm$1000 \kms. This result is consistent with previous studies (e.g., Fig. 3 
of \citealt{sul00a}; Fig. 2 of \citealt{baskin05}; Fig. 6 of 
\citealt{shang07}). 
Most recently, \citet{bonning07} also studied the velocity shift 
of \hbb\ using SDSS quasars, and our result is very similar to theirs; they 
fit their distribution of velocity shifts (their Fig. 1) using a Gaussian 
profile with a peak velocity of 100 \kms.  Our distribution of \mgii\ shifts is 
indistinguishable from that of \hbb.  It is symmetrical around 0, has a median 
value of 113 \kms\ and a maximum value of about $\pm$1000 \kms, and it is 
consistent with those given in, for example, \citet{richards02b} and 
\citet{shang07}. 

In order to compare \feii\ directly with \hbb, we also plot the distribution 
of the \feii\ velocity shift with respect to \hbb, as shown in Figure
\ref{fig-shift}{\it f}.  Again, most objects have redward shifts in \feii. 

\subsection{Fe Emission Widths}
\label{resultwidth}

\begin{figure*}
  \centering
  \includegraphics[angle=-90,width=0.9\textwidth]{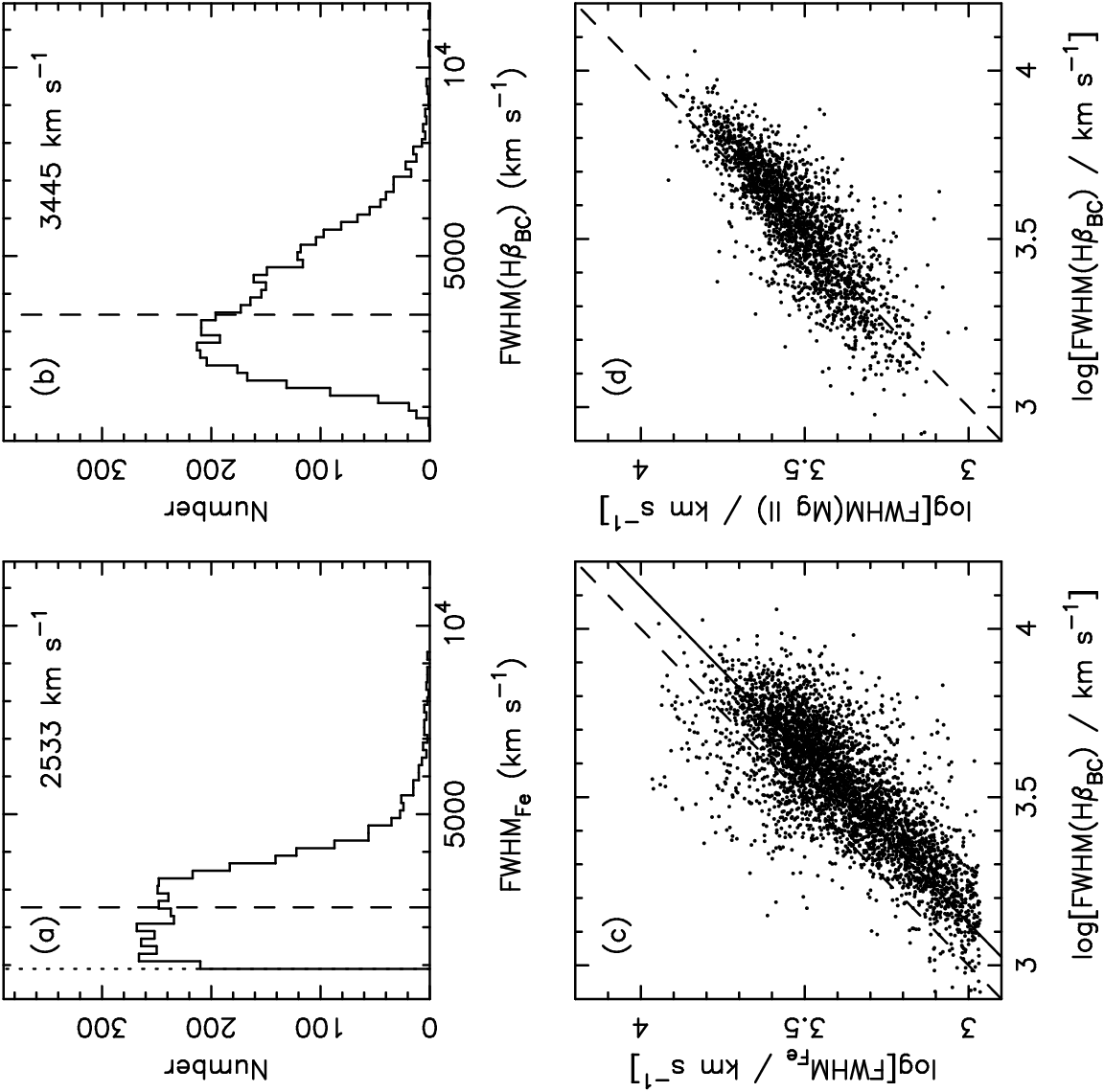}
  \caption{ Distribution of ({\it a}) \feii\ FWHM and ({\it b})
  \hbb\ FWHM. The number in the panel is the median FWHM, which is also marked 
  by the dashed line. \fwhmfe\ has a lower limit of 900 \kms\ (dotted line); 
  this is artificial because the I Zw 1 Fe template we used has a width of 900 
  \kms.  ({\it c}) Correlation between \fwhmfe\ and \fwhmhbb.  The dashed 
  diagonal line shows \fwhmfe\ = \fwhmhbb, and the solid line is \fwhmfe\ = 
  3/4 \fwhmhbb. ({\it d}) Correlation between \fwhmfe\ and \fwhmmg.  The 
  dashed diagonal denotes \fwhmhbb\ = \fwhmmg.}
  \label{fig-width}
\end{figure*}

Figures \ref{fig-width}{\it a}\ and \ref{fig-width}{\it b}\ show the 
distribution of \fwhmfe\ and \fwhmhbb, respectively. \fwhmfe\ has an artifical 
lower limit of 900 \kms, which is bounded by the I~Zw~1 \feii\ template.  The 
median value of \fwhmfe, 2533 \kms, is $\sim 0.74$ of that of \fwhmhbb\ (3445 
\kms).  Almost all the sources have \feii\ lines narrower than \hbb, and the 
majority have \fwhmfe\ $\approx$ 3/4 \fwhmhbb\ (Fig. \ref{fig-width}{\it c}). 
This result is contrary to the prevailing notion that \feii\ and \hbb\
have similar profiles and are emitted from the same region 
\citep[e.g.,][]{bg92}, but is consistent with some more recent studies of a 
few objects \citep[e.g.,][]{popovic07,matsuoka07}. 

We find that \fwhmhb\ and \fwhmmg\ are well correlated and roughly equal 
(Fig. \ref{fig-width}{\it d}), consistent with \citet{mclure02,salv07}. 
\citet{salv07} find that \fwhmhb\ tends to be larger than \fwhmmg\ for 
\fwhmhb\ $>$ 4000 \kms, an effect they attribute to an extensive red wing on 
\hb. This tendency can also be seen in our plot, and our composite spectra 
(Fig. \ref{fig-comhb}) do show that \hb\ tends to have a red asymmetry when 
broad.

\subsection{Correlations with Other Emission-line Parameters}
\label{resulte1}

\citet{bg92} used principal components analysis to study the correlations 
among various observed properties of nearby quasars, and found that most of
the variance in the optical spectra of quasars is connected with the inverse
correlation between \feii\ and \oiii\ strength. The soft X-ray photon index
($\Gamma_{\rm soft}$), the ratio of \feii\ to \hbb\ (\rfe), and the line width 
of \hbb\ [\fwhmhbb] correlate with each other \citep{boller96,wang96,laor97a}.
\citet{sul00a,sul00b,sul07} identified that \fwhmhbb, \rfe, $\Gamma_{\rm
soft}$, and the velocity shift at half maximum of the broad \civ\ line profile 
[\chalf] provide discrimination between different AGN types. So, for simplicity,
instead of using a formal principal components analysis to study the 
correlations between \feii\ and other parameters, we just study the 
correlations with \fwhmhbb\ and \rfe.

\begin{figure*}
  \centering
  \includegraphics[angle=-90,totalheight=0.48\textwidth]{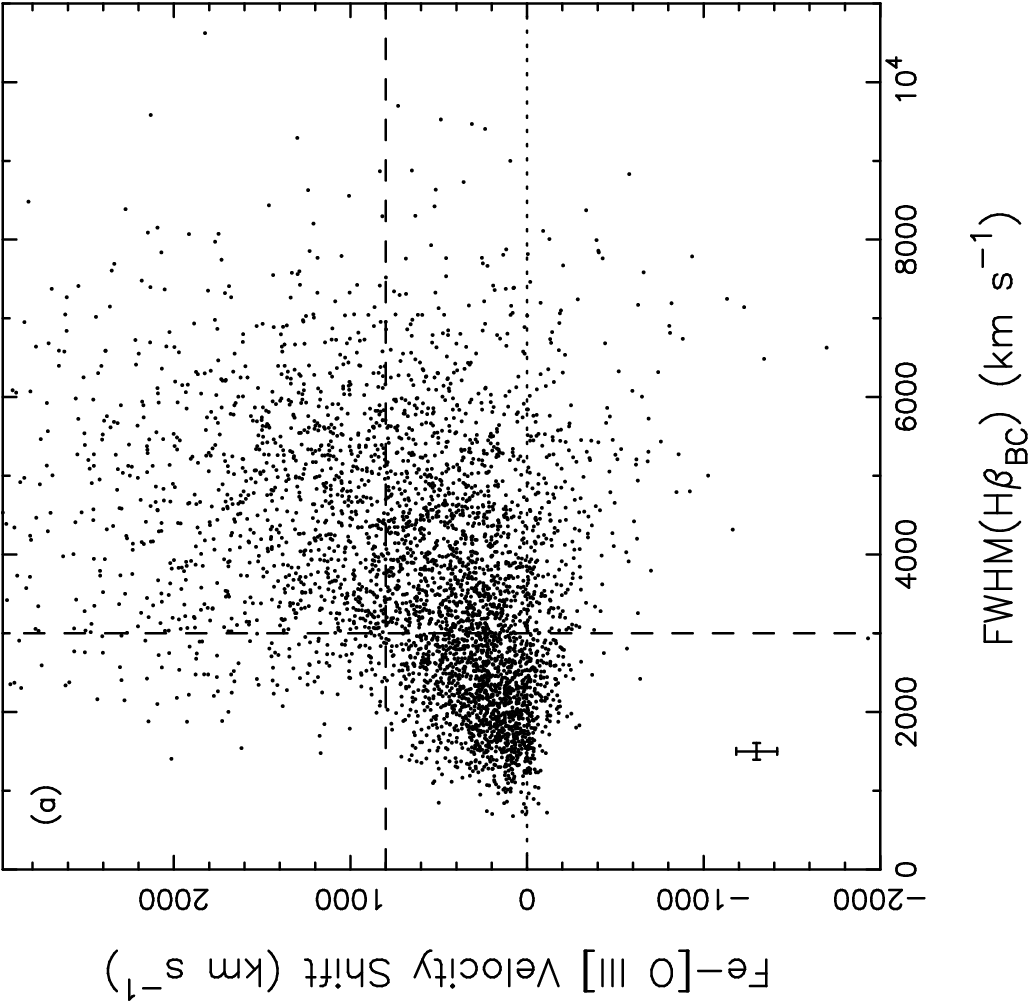}
  \hspace{0.3cm}\includegraphics[angle=-90,totalheight=0.48\textwidth]{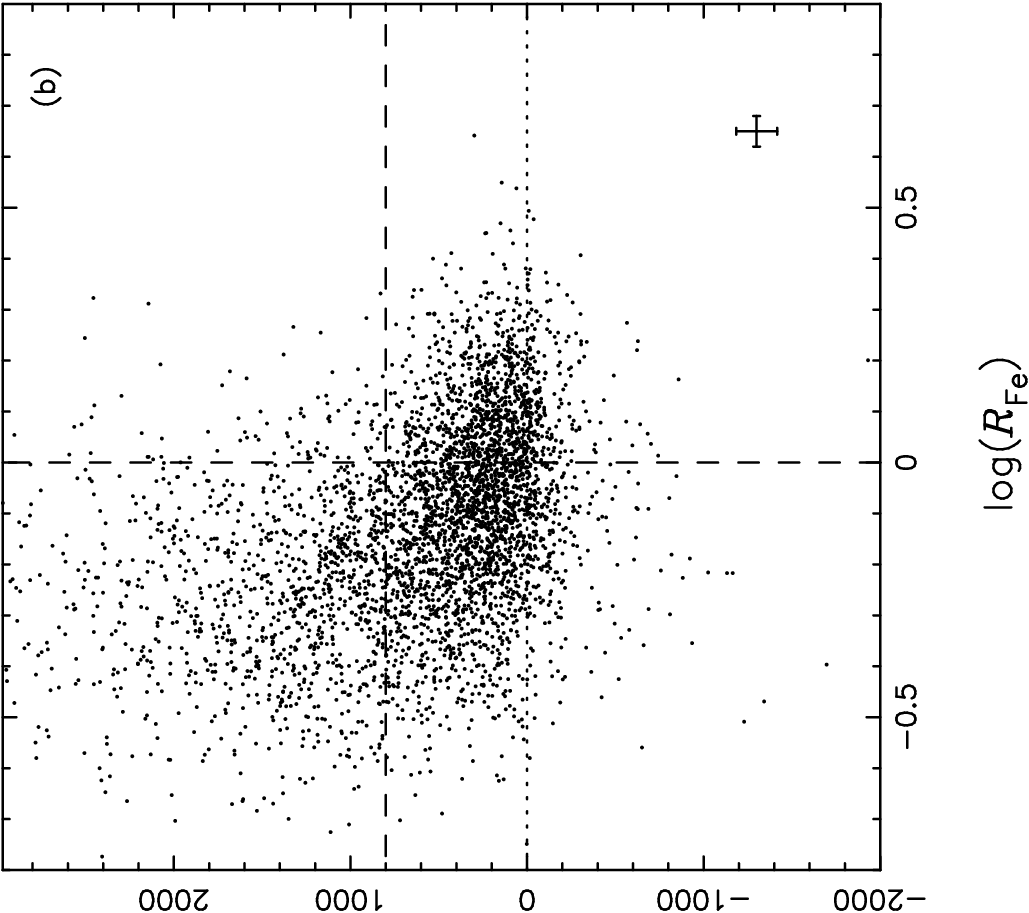}
  \caption{ Correlation diagram for ({\it left}) \vfe\ vs.
  \fwhmhbb\ and for ({\it right}) \vfe\ vs. \rfe. The crosses in the figure
  present the median values of the errors in both coordinates. See text for
  description of the vertical and horizontal lines.}
  \label{fig-e1shift}
\end{figure*}

The left panel of Figure \ref{fig-e1shift} shows the dependence of \vfe\ on
\fwhmhbb\ for our sample. The crosses in the figure present the median
values of the errors in both coordinates. Dividing the sample by a vertical
line of \fwhmhbb\ = 3000 \kms\ 
and a horizontal line of \vfe\ = 800 \kms, we find that almost
all sources with \fwhmhbb\ $<$ 3000 \kms\ have \vfe\ $<$ 800 \kms, and most 
sources with \vfe\ $>$ 800 \kms\ have \fwhmhbb\ $>$ 3000 \kms.  Sources with 
large \vfe\ and low \fwhmhbb\ are rare. There is also a large spread in \vfe\ 
for sources with \fwhmhbb\ $>$ 3000 \kms. Note that narrow-line Seyfert 1 
galaxies (NLS1s), defined by \fwhmhbb $<$ 2000 \kms\ \citep{oster85}, 
almost all have low \vfe\ in our sample.

Following \citet{netzer07}, we define \rfe\ as the line luminosity ratio of
\feii\ and \hbb,
\begin{equation}
  R_{\rm Fe}= \frac{L({\rm Fe} \textsc{ ii } \lambda 4570)}
  {L({\rm H}\beta_{\rm BC})},
\end{equation}
where $L({\rm Fe} \textsc{ ii } \lambda4570)$ is the luminosity of \feii\
emission between $\lambda$4434 and $\lambda$4684.  The right panel of
Figure \ref{fig-e1shift} shows the correlation diagram for \vfe\ versus \rfe. 
The vertical line is \rfe\ = 1. Most sources with \rfe\ $>$ 1 have small
\vfe. Sources with large \rfe\ and large \vfe\ are rare.
This correlation agrees with the correlation between \rfe\ and \er\ shown in
Figure 5 of \citet{netzer07}, considering the strong inverse correlation between
\vfe\ and \er\ as we find in \S \ref{resulter}, below.

The statistical connection between \vfe\ and \fwhmhbb\ or \rfe\ is similar to
those shown in Figures 2 and 9 of \citet{bg92}. This suggests that \vfe\ can
also provide useful empirical discrimination between different types of AGNs 
from optical spectra.

\subsection{The Physical Driver of \vfe}
\label{resulter}
The Eddington ratio \er\ is often suggested to be the main physical driver of 
the spectral diversity in quasars
\citep{sul00a,sul00b,marzi01,boroson02,marzi03a}, and \mbh\ is argued to be an
important determinant of radio-loudness
(\citealt{laor00,boroson02,mclure04}, but see \citealt{ho02}). This section 
investigates attempts to determine which physical variable is the main driver 
of variations in \vfe.

Central black hole masses can be estimated from empirical relations 
derived from reverberation mapping. We derived the BH mass and \er\ using
the relation calibrated by \citet{mcgill08},
\begin{eqnarray} 
  {\rm log} \left(\frac{M_{\rm BH}}{M_\odot}\right) & = & 7.383 + 2~
  {\rm log} \left(\frac{\sigma_{{\rm H}\beta}}{1000~{\rm km~s^{-1}}}\right)
  \nonumber\\
  & & + 0.69~
  {\rm log} \left(\frac{\lambda L_{5100}}{10^{44}~{\rm ergs~s^{-1}}}\right)
  \label{equ-m}
\end{eqnarray}
(their Table 3; we use the factors for $L_{5100,t}$ and
$\sigma_{H\beta}$).
We estimate the bolometric luminosity using 
$L_{\rm bol}=9\lambda L_{5100}$ \citep{kaspi00}.

\begin{figure*}
  \centering
  \includegraphics[angle=-90,width=0.45\textwidth]{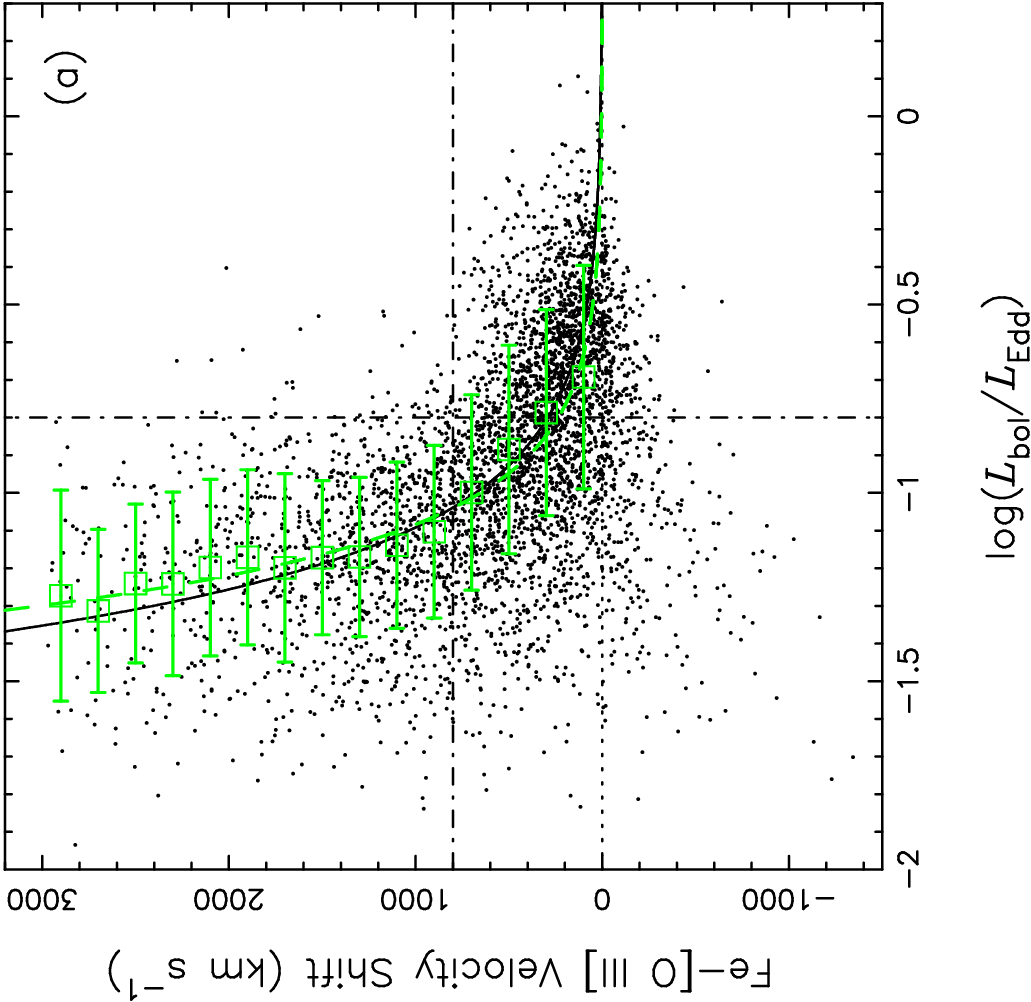}
  \hspace{0.4cm}\includegraphics[angle=-90,width=0.45\textwidth]{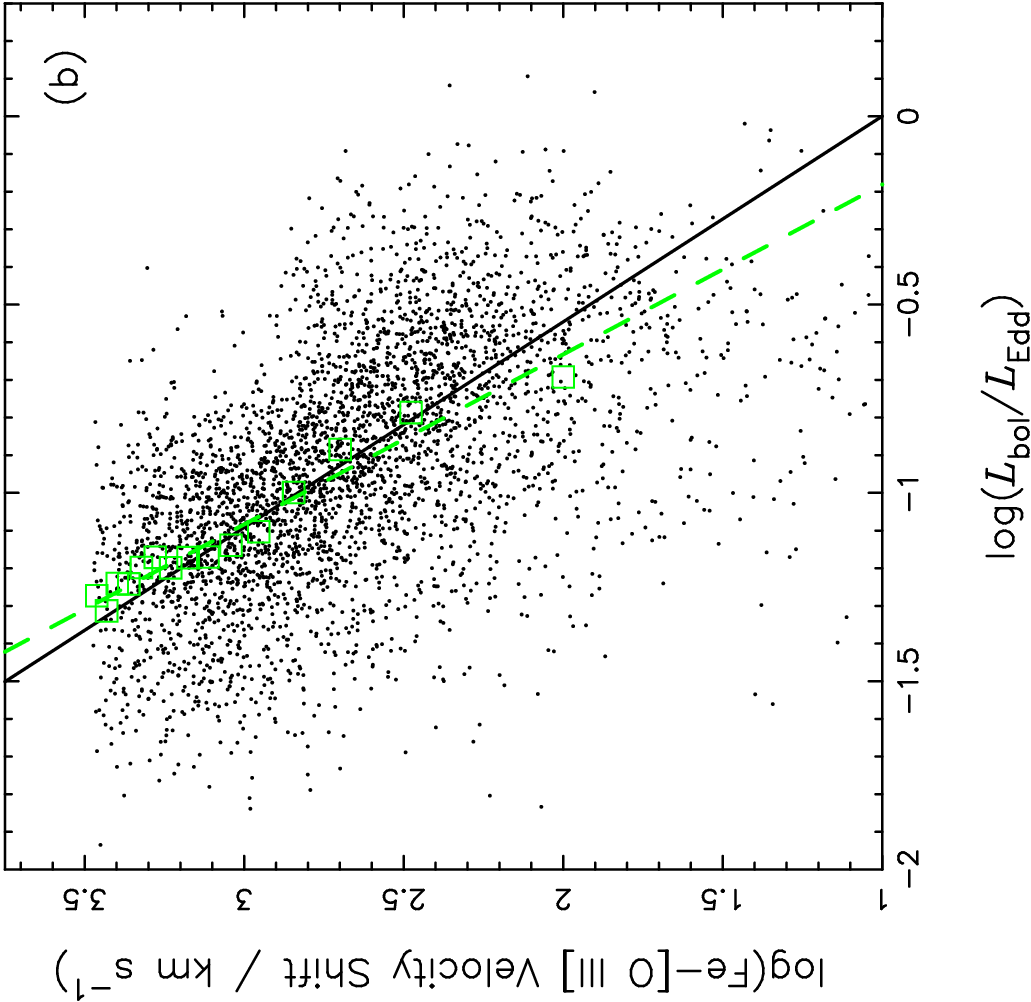}
  \caption{ Correlation of \feii\ velocity shift vs. Eddington 
  ratio, with the \feii\ velocity shift plotted ({\it a}) linearly and 
  ({\it b}) logarithmically.  See text explanation of the dot-dashed lines. 
  The solid lines is the fit to data in Eq. (\ref{equ-cor}).  The 
  green squares represent the mean log(\er) in bins of 
  $\Delta v_{\rm Fe}=200~{\rm km~s^{-1}}$, and the error bars represent the 
  standard deviations.  The green dashed line shows the fit to the binned data. 
  }
  \label{fig-eddshift}
\end{figure*}

Figure \ref{fig-eddshift}{\it a}\ shows \vfe\ as a function of \er\ for our sample.
The vertical dot-dashed line is log(\er) = $-0.8$, and the horizontal dot-dashed
line is \vfe\ = 800 \kms.  We find that almost all sources with 
log(\er) $>$ $-0.8$  have \vfe\ $<$ 800 \kms, and most sources with \vfe\ 
$>$ 800 \kms\ have log(\er) $<$ $-0.8$.  There are very few sources with large 
\vfe\ and large \er. The error in \er\ is roughly 0.3 dex, which is
dominated by the systematical error in estimating the BH mass using
empirical relations \citep{mcgill08}.

The logarithmic form of this diagram is plotted in Figure 
\ref{fig-eddshift}{\it b}, excluding the 481 sources with negative \vfe. We 
find a strong inverse correlation between \vfe\ and \er.  The fit taken into
account the uncertainties in both quantities yields the solid line:
\begin{equation}
  \log v_{\rm Fe}=(1.00\pm0.05)
  -(1.83\pm0.05)~{\rm log}(L_{\rm bol}/L_{\rm Edd}). 
  \label{equ-cor}
\end{equation}
Pearson's correlation coefficient $r_{\rm P}$ is $-0.53$, and the probability 
$P$ of a chance correlation $< 1\times10^{-5}$.  
Note that below an \vfe\ of about 150 \kms, the scatter to the fitted line
increases. As mentioned in \S \ref{mcsim}, the input value of \vvfe\ affects
the measurements little; this means that the errors of \vfe\ will not decrease
with \vfe. Thus, for sources with small \vfe, the fractional error on \vfe\
will be large.  This causes the scatter described before.  The squares show
the mean values of log(\er) in bins of $\Delta v_{\rm Fe}=200~{\rm
km~s^{-1}}$; the error bars in panel ({\it a}) show the standard deviations.
The dashed line shows the fit to the binned values of log(\er): 
\begin{equation}
  \log v_{\rm Fe}=(0.60\pm1.40)
  -(2.22\pm0.99)~{\rm log}(L_{\rm bol}/L_{\rm Edd})~,
  \label{equ-cor2}
\end{equation}
In this case, $r_{\rm P}=-0.98$ and $P < 1\times10^{-5}$. The above analysis
indicates that \vfe\ depends strongly on \er; $v_{\rm Fe}~\propto~(L_{\rm
bol}/L_{\rm Edd})^\gamma$, with $\gamma\approx -2$: the larger the Eddington 
ratio, the lower the velocity shift \vfe.  Plotting \vfe\ as an function of 
\mbh\ and \lbol\ (not shown) reveals that neither of these two variables is as 
important as \er.  The Eddington ratio is the main physical driver for \vfe.
This result provides a strong constraint on theoretical models of the \feii\ 
emission region.

\subsection{Correlations with Radio and X-ray Properties}
\label{resultother}

\begin{figure}
  \centering
  \includegraphics[angle=-90,width=0.45\textwidth]{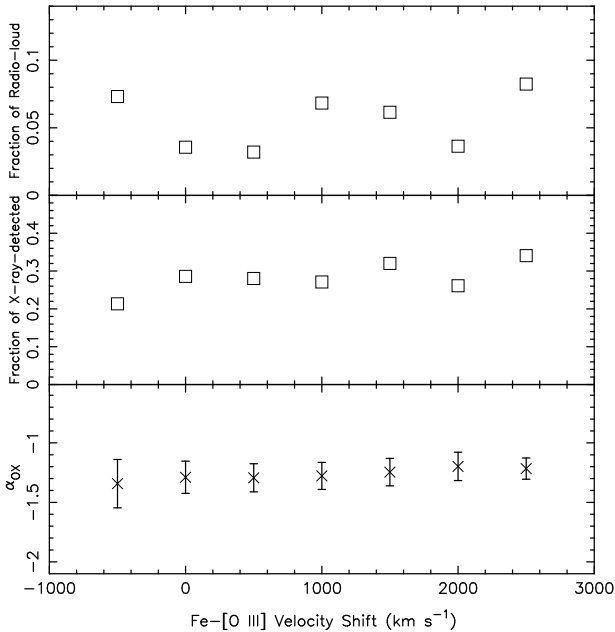}
  \caption{ ({\it Top}) The fraction of radio-loud quasars in
  bins of different \vfe.  
  ({\it Middle}) The fraction of X-ray-detected quasars.
  ({\it Bottom}) \aox\ in bins of different \vfe. }
  \label{fig-frac}
\end{figure}

In an effort to understand the physical origin of the \feii\ velocity shift, we 
examine whether \vfe\ correlates with radio and X-ray emission. 
\citet{richards02b} conducted a similar investigation in their analysis of 
velocity shifts for the \civ\ line.  Figure \ref{fig-frac} (upper panel) 
plots the fraction of radio-loud quasars in bins of different \vfe.  We define 
the radio-loudness parameter as $R={\rm log}(L_{\rm 6 cm}/L_B)$, where 
$L_{\rm 6 cm}$ and $L_B$ are the observed luminosities at 6 cm and 4400 \ang.
We use the FIRST \citep{becker95} peak flux densities at 20 cm in Table 2 of 
\citet{schneider07} to calculate the radio luminosity, assuming a radio 
spectral index $\alpha_{\rm r} = -0.5$.  The optical luminosities are 
calculated from the power-law continuum we fitted. We classify the sources 
with $R > 1$ as radio-loud.  There are a total of 165 radio-loud quasars
out of 3750 sources in the present sample within the FIRST survey area. Note 
that the percentage of radio-loud quasars in our sample is $165 / 3750 = 
4.4\%$, slightly lower than the 6.7\% found by \citet{mclure04}.  The
reason is probably that our sample is biased toward quasars with high \er\
(by our \feii\ EW cut), since $R$ decreases with increasing
\er\ \citep{ho02,greene06a}. 
The rightmost bin
have the largest radio-loud fraction, in which there are 7 radio-loud
quasars out of 85, a fraction $\sim$2 times higher than average. The 
cumulative Poisson probability ($P$) of getting 7 or more objects 
out of 85 is 
$8.54\times10^{-2}$ \citep{gehrels86} when the average is 3.74 ($165 / 3750
\times 85$), no more than a $3\,\sigma$ significance ($1-P$ $<$ 99.73\%). 
Thus there is no clear trend of radio-loud fraction with \vfe. 

Next, we evaluate the fraction of X-ray-detected quasars as a function of 
\vfe\ (middle panel of Fig. \ref{fig-frac}). We use the data from
\citet{schneider07}, who provide the X-ray full-band count rate from the {\it
ROSAT}\ All-Sky Survey Bright \citep{voges99} and Faint \citep{voges00} 
sources catalogs.  No obvious trend is apparent.  Finally, we test for 
possible dependence of \vfe\ on the optical-to-X-ray spectral index 
$\alpha_{\rm OX} \equiv -0.3838~{\rm log} (L_{\rm 2500}/L_{\rm 2~keV})$, where 
$L_{\rm 2500}$ is the specific luminosity at 2500 \ang\
calculated using the power-law continuum we measured and $L_{\rm 2~keV}$ is the
specific luminosity at 2 keV derived from the {\it ROSAT}\
count rate using {\tt PIMMS} \citep{mukai93} assuming a power-law model with
photon index of 2. We see no obvious trend between \aox\ and \vfe\ either
(bottom panel of Fig. \ref{fig-frac}).

\subsection{Composite Spectra}
\label{composite}
In order to get a visual impression of the correlations between \vfe\ and
other emission-line properties, we create a set of five composite spectra by
combining quasars in bins of different \vfe. We divide our sources into five 
subsamples, covering the following velocity ranges: $-$250 to 250 \kms\ (A, 
1350 objects), 250 to 750 \kms\ (B, 1362 objects), 750 to 1250 \kms\ (C, 590 
objects), 1250 to 1750 \kms\ (D, 332 objects), and 1750 to 2250 \kms\ (E, 180 
objects).  The composite spectra are generated following the procedure of
\citet{vb01}. The spectra of quasars in each subsample are deredshifted using 
the redshifts determined from \oiii\ and then normalized to unity average flux 
density over the rest wavelength interval 5090--5110 \ang. We generate the 
composites using the geometric mean, which is appropriate for quasars 
with power-law spectra because the geometric mean will result in a power law 
with the mean spectral index \citep{vb01}. 

\begin{figure*}
  \centering
  \includegraphics[angle=-90,width=0.9\textwidth]{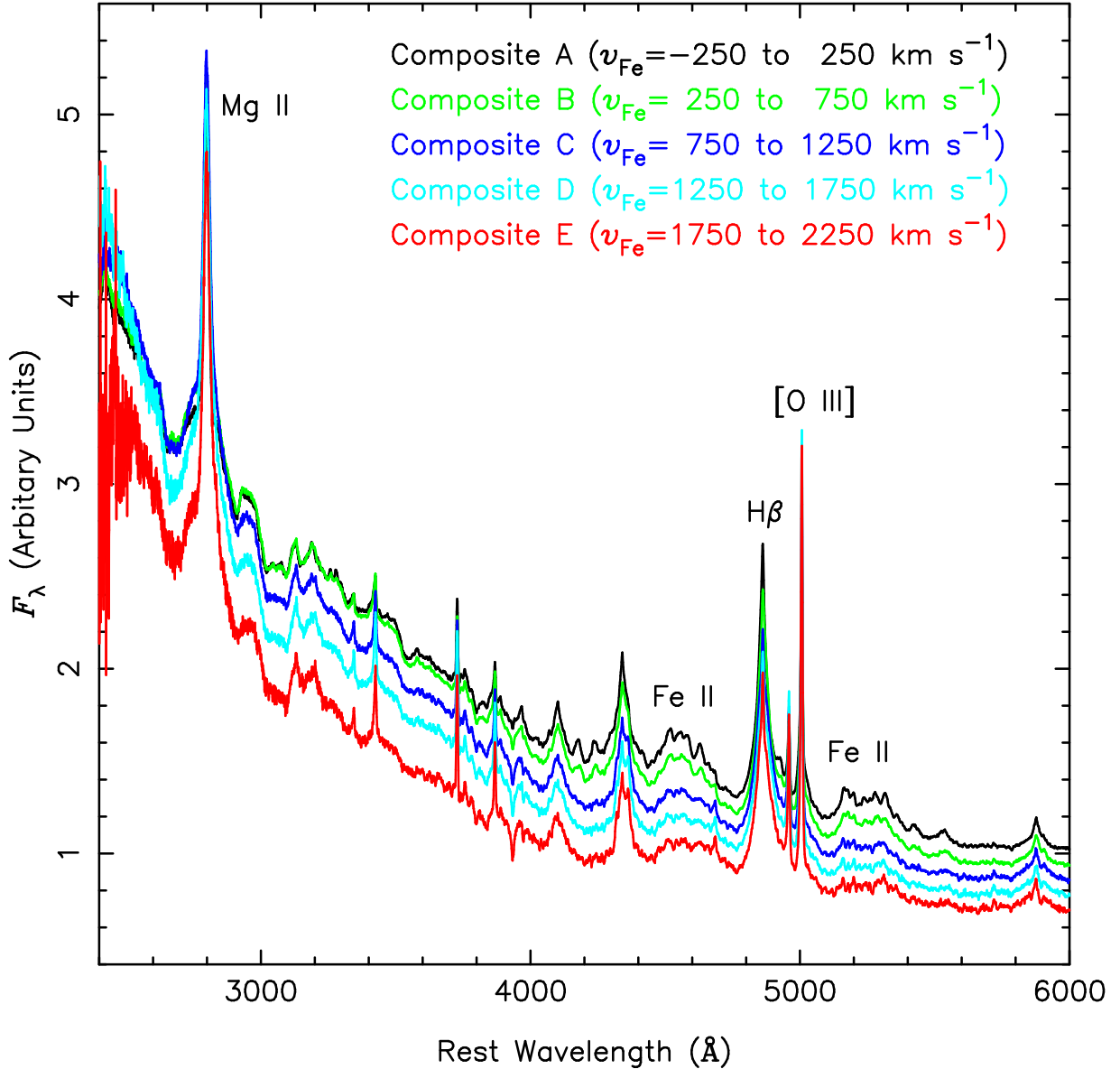}
  \caption{ Composite spectra of quasars in bins of different
  \vfe. See text for details on generating the composite spectra. We shifted 
  the spectra vertically for clarity.  Major emission lines are
  indicated; see \citet{vb01} for a complete identification of the emission
  lines.}
  \label{fig-com}
\end{figure*}

Figure \ref{fig-com} shows the composite spectra of the five subsamples
described above. We plot the spectra in different colors and shifted them
slightly vertically for clarity.  The spectra are arranged so that, from top
to bottom, the velocity shifts of the \feii\ emission increase.  We find that 
\fwhmhbb\ increases while the \feii\ flux decreases, consistent with the
correlations seen in \S \ref{resulte1}. Inspecting the continuum slope,
composite A shows a redder continuum than the rest (this can be seen most 
easily
when comparing composites A with B). This suggests that sources with low \vfe,
which from our analysis are often accompanied by high \er\ and strong 
\feii\ emission, tend to have redder UV-optical continua. This pattern is 
reminiscent of that seen by \citet{constantin03}, who found that the
UV-optical continuum of NLS1s, which have high \er\ and strong \feii, is redder
than that of regular AGNs. This is unexpected from standard accretion disk 
models, since a higher \er\ generally produces a hotter disk and thus a bluer
continuum \citep[e.g.,][and references therein]{hubeny00}.

Note that it is difficult to see the shifts of the \feii\ emission clearly in 
the composite spectra.  For example, the peaks between $\sim 5150$ and $5250$ 
\AA\ appear to remain unshifted. The reason is as follows.  On the one hand, 
since both the shifts and widths of the \feii\ emission have a large scatter 
in each subsample, stacking the spectra effectively smooths the \feii\ 
emission. On the other hand, there are many narrow emission lines (e.g., line
system N3 in \citealt{veron04}) that are weak in a single spectrum and have
nearly no shift.  Stacking the spectra enhances these narrow features.  

\begin{figure}
  \centering
  \includegraphics[angle=-90,width=0.45\textwidth]{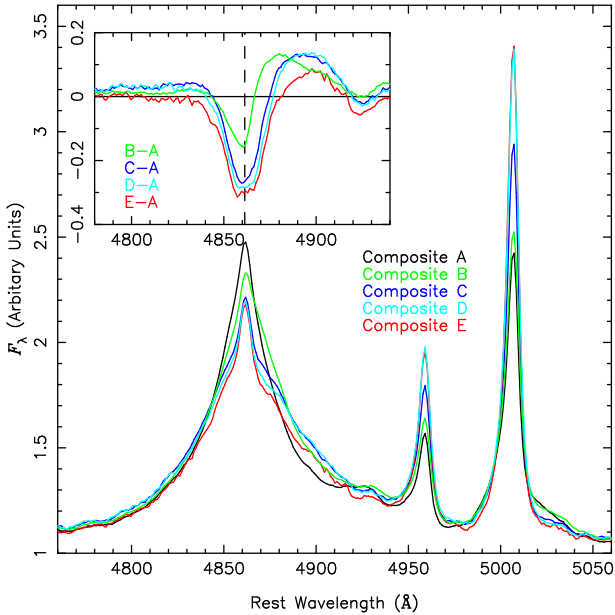}
  \caption{ \hb\ and \oiii\ region of the composite
  spectra. The inserted plot shows the difference spectra.
  The vertical dashed line is drawn at 4861 \ang. 
  Note the excess on the red wing of \hb.
  }
  \label{fig-comhb}
\end{figure}

Another interesting phenomenon is that the wings of the \hb\ profile become 
progressively more (red) asymmetric when \vfe\ increases (Fig. \ref{fig-comhb}).
The inserted plot shows the difference spectra between composites B through 
E, using composite A as reference, to emphasize the profile changes.
Note the systematic migration of the red excess as \vfe\ increases.
We do find some individual sources, similar to OQ 208 \citep{marzi93},
whose redshifted \hb\ component seems to be associated with \feii\ emission 
(\S \ref{shifthb}).  The asymmetry in \hb\ has been studied by many authors.
\citet{bg92} found that there are \hb\ red asymmetries at small \rfe.
\citet{marzi96} found that radio-loud AGNs show predominantly redshifted and
red asymmetric \hb\ profiles: the larger the shift, the broader the line.
Recently, \citet{netzer07} investigated the fractional luminosity of the red
part of the \hb\ line and found that it has a tendency to increase with
decreasing \feii/\hb\ (see their Table 3). Considering that \vfe\
inversely correlates with \rfe, as shown in the right panel of
Figure \ref{fig-e1shift}, the finding here is consistent with theirs.  One
possible interpretation, discussed in \S \ref{feorig}, is that the \hb\ excess 
emerges from the same region that produces \feii.  A systematic
study of this issue will be carried out in a future paper.

\section{Discussion}
\label{discuss}

\subsection{Redshifted \hb\ Component: OQ 208-like Quasars}
\label{shifthb}
In our sample, we find a class of sources whose \hb\ profile can be fitted 
well including an additional, substantially redshifted Gaussian 
component. Usually this additional component has a velocity width intermediate
between that of the broad and narrow components. We call this the intermediate
component.  An interesting and possibly highly significant fact is that the 
width and velocity shift of this additional \hb\ component is consistent with 
those of the \feii\ emission. A prototype of this kind of sources is OQ 208 
(Mrk 668), which was first studied in detail by \citet{marzi93}. They pointed 
out that in OQ 208 the \feii\ lines at 4924 and 5018 \ang\ have the same peak 
displacement as the red peak of \hb. 

\begin{figure*}
  \centering
  \noindent\includegraphics[angle=-90,width=0.9\textwidth]{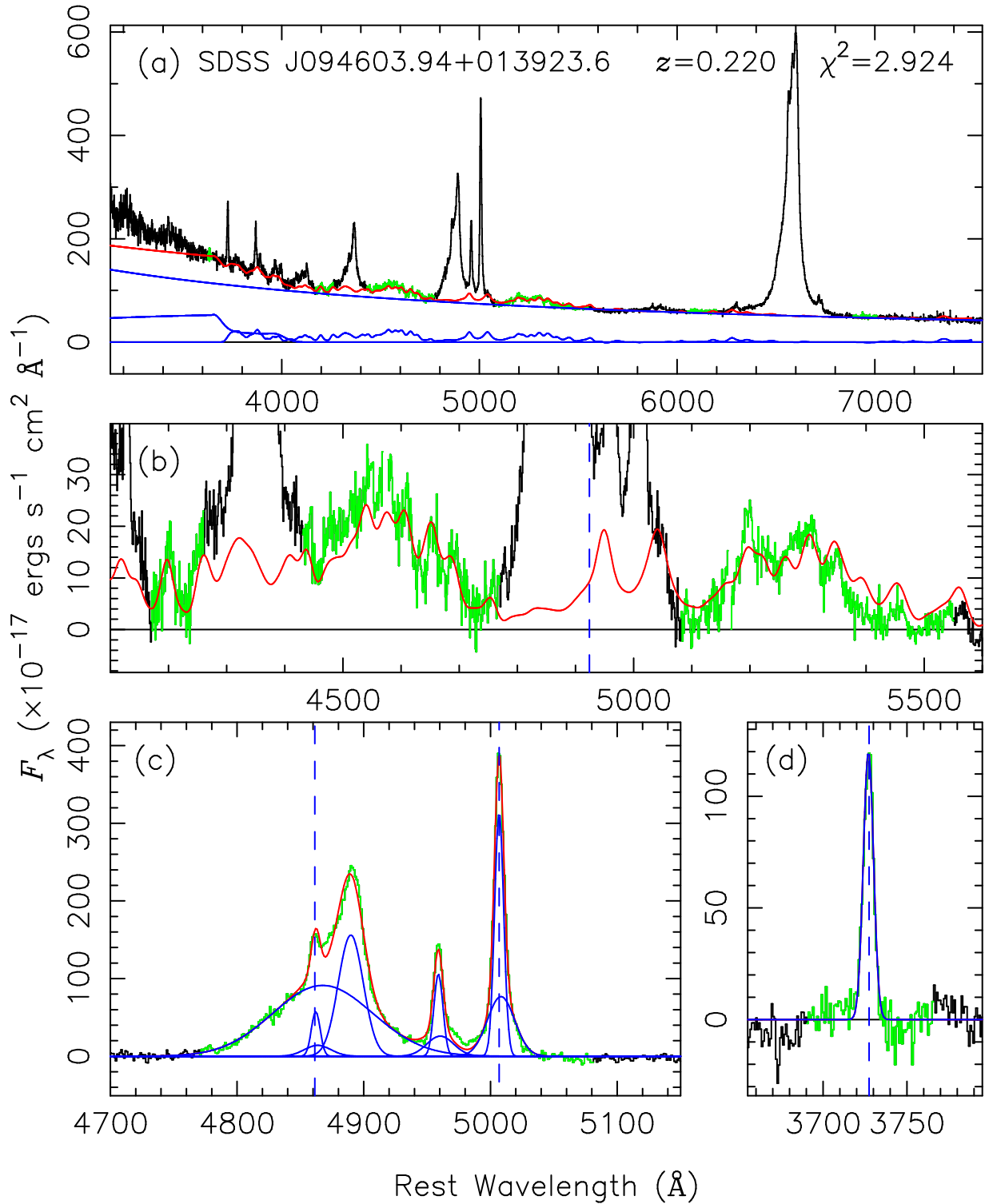}
  \caption{ Spectrum of SDSS J094603.94+013923.6, an example of a 
  OQ 208-like source. ({\it a}) Continuum decomposition. ({\it b})
  Fe model and the observed spectrum after subtracting the power law and Balmer
  continuum.  The spectra and the marks plotted in panel ({\it a}) and ({\it
  b}) are as the same as those in the top and bottom panels of Fig.
  \ref{fig-conti1}.  ({\it c}) Fitting of \hb\ and \oiii\ emission lines. The 
  multiple Gaussian components are in blue, and the sum of them is in red. The 
  two blue dashed lines mark the rest-frame wavelength of \hb\ and \oiii\ 
  $\lambda$5007. ({\it d}) Fitting of \oii\ emission line.  The blue dashed 
  line marks the rest-frame wavelength of \oii\ $\lambda$3727.  Note that the 
  redshifts and widths of \oii, \oiii, and \hbn\ are almost the same.  Note 
  also that we added an additional redshifted Gaussian line to fit \hb. The
  width and velocity shift of this redshifted \hb\ component are consistent
  with those of \feii\ emission. See text and Table \ref{tab-redhb} for more
  details.} 
  \label{fig-redhb}
\end{figure*}

Figure \ref{fig-redhb} shows the spectrum of SDSS J094603.94+013923.6, an 
example of a OQ 208-like source.  Panel ({\it a}) shows the continuum 
decomposition, following the same convention as used in the top panel of 
Figure \ref{fig-conti1}. Panel ({\it b}) shows the emission-line spectrum after 
continuum subtraction.  The red solid line is our \feii\ model.  The blue 
dashed line marks the position of the peak of \feii\ $\lambda$4924 at zero 
velocity shift. The shift of the \feii\ spectrum is obvious.  Panel ({\it c})
illustrates the detailed fitting of the \hb\ and \oiii\ emission lines.  The 
profile of \hbn\ is fixed to that of \oiii.  The two blue dashed lines mark 
the rest-frame wavelength of \hb\ and \oiii\ $\lambda$5007; both \hbn\ and 
\oiii\ share the same velocity.  A prominent, intermediate-width redshifted 
component is clearly required to fit \hb.  Panel ({\it d}) shows the fit for 
\oii\ $\lambda$3727, which also has the same redshift as \oiii.  This 
means that this source is not an \oiii\ ``blue outlier'' \citep{boroson05}.
Table \ref{tab-redhb} lists the FWHMs and velocity shifts of \feii, \oiii,
\oii, and each component of \hb. The velocity shifts are all with respect to 
\oiii. The width and velocity shift of the redshifted \hb\ component are 
consistent with those of \feii\ emission.  The emission-line spectrum of 
SDSS J094603.94+013923.6 can be divided into three systems: (1) \hb, \oiii, 
and \oii\ emission lines with narrow (FWHM $\approx 500$ \kms) widths and no 
velocity shift; (2) a normal ``broad'' \hb\ component with FWHM $\approx 5700$ 
\kms, approximately at rest with respect to the narrow lines; and (3) \feii\ 
emission and \hb\ of intermediate width (FWHM $\approx 1500$ \kms) 
redshifted by $\sim 1500-1700$ \kms.

\begin{deluxetable*}{ccccccc}
  \tablewidth{0pt}
  \tablecolumns{7}
  \tablecaption{Emission-line Properties of SDSS J094603.94+013923.6
  \label{tab-redhb}}
  \tablehead{
  \colhead{} & \colhead{Fe} & \multicolumn{3}{c}{\hb} &
  \colhead{\oiii\tablenotemark{a}}
  & \colhead{\oii} \\
  \cline{3-5} 
  \colhead{} & \colhead{} & \colhead{Redshifted} & \colhead{Broad} &
  \colhead{Narrow} & \colhead{} & \colhead{}
  }
  \startdata
  FWHM (\kms)  & 1543(55) & 1428(18) & 5730(58) & \nodata\tablenotemark{b} &
  505(6) & 568(18) \\
  Velocity shift (\kms) & 1533(24) & 1756(7)  &
  373(25)  & 49(10) & \nodata\tablenotemark{c} & $-$22(8) 
  \enddata
  \tablecomments{The number in parenthesis is the error.}
  \tablenotetext{a}{Refers to the line core component used to derive
  the redshift.}
  \tablenotetext{b}{The width of the \hbn\ is fixed to the width of \oiii.}
  \tablenotetext{c}{The redshift of \oiii\ is used as the systemic redshift.}
\end{deluxetable*}

OQ 208-like sources can offer an unique view to understand the origin of
\feii\ emission and the structure of the BLR.  A detailed study of this class 
of objects is beyond the scope of this paper.  We will analyze a sample of 
such sources discovered in our work in a forthcoming publication.  
For the purposes of the present discussion, we simply note: \feii\ emission 
is not associated with the conventional broad component of \hb\ but instead
originates from the same region emitting the intermediate component of \hb. 
This 
is consistent with our findings in \S \ref{resultshift} that the width of 
\feii\ is systematically narrower than that of \hbb.

\subsection{Where is the \feii\ Emission Region?}
\label{feorig}

This paper has demonstrated that \feii\ emission most likely does not 
originate from the same location that produces the broad component of \hb.
The evidence comes from the systematic redshifts and narrower line widths 
of \feii\ compared to \hbb.  As discussed below, the simplest interpretation 
is that \feii\ emission originates from an inflow that is located at the outer 
parts of the BLR.  The \hbb\ emission line itself shows no systematic velocity 
shift, and any velocity shift itself is also small. This suggests that \hbb\ 
emission region is well virialized and that the width of \hbb\ is dominated by 
gravity, making this line suitable for estimating BH masses.  

Our finding that \fwhmfe\ $\approx 3/4$ \fwhmhbb\ suggests that the \feii\
emission region is farther from the central BH than the \hbb\ emission region.
If the \feii\ emission region is also virialized, so that $R\propto v^{-2}$, 
then it is about 2 times farther from the center than the \hbb\ region.  On 
the other hand, the systematic redshift of \feii\ indicates that the 
assumption of virialization may be incorrect.

The redward shift of \feii\ and the inverse correlation between \vfe\ and \er\
favor a scenario in which \feii\ emission emerges from an inflow. To explain 
the systematic redshift, the inflow on the back side of the accretion disk must
be obscured by the accretion disk and the torus, so that we can only observe 
the redshifted part on the nearer side.  To explain the inverse correlation 
between \vfe\ and \er, we speculate that the inflow is driven by gravity 
toward the center and decelerated by the radiation pressure.  An increase in 
\er\ would enhance the radiation pressure and lead to a decrease of the
inward velocity of the inflow. 

Previous studies of the \civ\ line have shown that it tends to be 
systematically blueshifted 
\citep[e.g.,][]{gastell82,wilkes84,marzi96,sul00a,richards02b}.  
This finding has led to suggestions
that high-ionization lines and low-ionization lines
originate from distinct regions \citep[see also the result of reverberation
mapping][]{peterson99}. High-ionization lines such as \civ\ may be emitted 
from some kind of outflowing disk wind 
\citep[e.g.,][and references therein]{leighly04a,leighly04b,baskin05,shang07}, 
whereas low-ionization lines such as \hb\ are anchored to a more 
disk-like configuration.  This paper adds an additional element to this 
picture.  We suggest that in addition to a disk and a wind, the BLR has yet 
another component, one associated with inflowing material that produces 
the \feii\ emission.  Interestingly, \citet{welsh07} recently reanalyzed the 
spectral variability of the well-studied Seyfert 1 galaxy NGC 5548 and 
suggested, based on the differential lag between the red and blue wings of the 
\hb\ profile, that the BLR in this object contains an inflowing component.
Such an inflow is likely to develop from the inner edge of the dusty torus,
which may connect with the accretion disk. If this picture is correct, the 
redshifted \feii\ emission can be used as a probe of the transition region
from the dusty torus to the BLR or accretion disk.  
More detailed theoretical modelling and observations of the \feii\ emission 
region are clearly required to test this picture.

\section{Summary}
\label{summary}

Using a large sample of quasars selected from SDSS, we have studied the 
properties of their optical \feii\ emission, especially their velocity profile
and velocity shift, with the goal of understanding the origin of the nature 
of the \feii-emitting region.  This was accomplished using an improved iron 
template-fitting method, whose reliability has been tested using extensive 
simulations. 

Our findings can be summarized as follows:

1. The majority of quasars show redshifted \feii\ emission with respect to the 
systemic velocity of narrow-line region (as traced by \oiii\ $\lambda$5007) 
or of the conventional broad-line region (as traced by \hb).  The shift is
typically $\sim 400$~\kms\ and can be as large as $\sim 2000$ \kms. By 
contrast, neither the broad \hb\ nor \mgii\ lines show a systematic velocity
shift.

2. The velocity width of \feii\ is systematically narrower than that of the 
broad component of \hb.  On average, \fwhmfe\ $\approx$ 3/4 \fwhmhbb.

3. The velocity shift of \feii\ increases with decreasing Eddington ratio, 
decreasing Fe strength, and increasing \hb\ line width.  No clear trends 
with radio or X-ray properties can be discerned.

4. Composite spectra reveal that objects with large \feii\ velocity shifts 
have a tendency to exhibit asymmetric \hb\ profiles with an excess red
wing.  This phenomenon is particularly notable in a subclass of objects 
that resembles the prototype OQ 208.  The \hb\ profile of OQ 208-like sources 
show a redshifted component of intermediate width that closely resembles 
the \feii\ emission.

5. Our results strongly indicate that \feii\ emission does not originate from 
the same region of the BLR that produces the ``traditional'' broad 
component of \hb.  Instead, we suggest that \feii\ is associated with the 
intermediate-width \hb\ component, both tracing an inflowing component of 
the BLR.

\acknowledgments
We thank the referee, Kirk Korista, for his careful, detailed comments and
suggestions that helped to improve the paper.
J.M.W. is very grateful to A. Laor for useful comments.  We thank T. A.
Boroson and M. Vestergaard for the I Zw 1 Fe templates and their 
suggestions on the spectral fitting.  We appreciate extensive discussions
among the members of IHEP AGN group. The research is supported by NSFC and CAS
via NSFC-10325313, 10733010 and 10521001, and KJCX2-YW-T03, respectively.
This paper has used data from SDSS, FIRST, and {\it ROSAT}.  We are grateful
to the SDSS, FIRST, and {\it ROSAT}\ collaborations for their effort
devoted to conducting the surveys and providing the data to the public.
The FIRST survey is supported in part under the auspices of the Department of
Energy by Lawrence Livermore National Laboratory under contract W-7405-ENG-48
and the Institute for Geophysics and Planetary Physics.
The ROSAT Project is supported by the Bundesministerium f\"ur Bildung und
Forschung (BMBF/DLR) and the Max-Planck-Gesellschaft (MPG).
Funding for SDSS and SDSS-II has been provided
by the Alfred P. Sloan Foundation, the Participating Institutions, the
National Science Foundation, the U.S. Department of Energy, the National
Aeronautics and Space Administration, the Japanese Monbukagakusho, and the Max
Planck Society, and the Higher Education Funding Council for England.  The
SDSS Web site is http://www.sdss.org/. The SDSS is managed by the Astrophysical
Research Consortium (ARC) for the Participating Institutions.  The
Participating Institutions are the American Museum of Natural History,
Astrophysical Institute Potsdam, University of Basel, University of Cambridge,
Case Western Reserve University, The University of Chicago, Drexel University,
Fermilab, the Institute for Advanced Study, the Japan Participation Group, The
Johns Hopkins University, the Joint Institute for Nuclear Astrophysics, the
Kavli Institute for Particle Astrophysics and Cosmology, the Korean Scientist
Group, the Chinese Academy of Sciences (LAMOST), Los Alamos National
Laboratory, the Max-Planck-Institute for Astronomy (MPIA), the
Max-Planck-Institute for Astrophysics (MPA), New Mexico State University, Ohio
State University, University of Pittsburgh, University of Portsmouth,
Princeton University, the United States Naval Observatory, and the University
of Washington.

\end{document}